\documentclass[useAMS,usegraphicx,usenatbib]{mn2e} 
\bibpunct{(}{)}{;}{a}{}{,}
\usepackage{url,makeidx,xcolor,amsmath,amssymb,times,graphicx}
\usepackage{enumerate} 
\graphicspath{{img/},{img_color/}}
\DeclareGraphicsExtensions{,.jpeg,.png,.eps,.pdf}

\newcommand\gaia{\textit{Gaia}}

\newcommand\electron{\ensuremath{\mathrm{e}^{-}}}
\newcommand\cemga{{\sc CEMGA}}

\newcommand\trapsigma{\ensuremath{\sigma_\mathrm{t}}}

\title[Digging supplementary buried channels]{Digging supplementary buried channels:\\
investigating the notch architecture within the CCD pixels on ESA's {\gaia} satellite}

\author[G. M. Seabroke et al.]{
G.M.~Seabroke$^{1,2}$\thanks{E-mail: gms@mssl.ucl.ac.uk},
T.~Prod'homme$^{3,4}$,
N.J.~Murray$^{2}$,
C.~Crowley$^{5}$,
G.~Hopkinson$^{\dagger,6}$,
\newauthor
A.G.A.~Brown$^{4}$,
R.~Kohley$^{5}$,
A.~Holland$^{2}$\\
$^1$Mullard Space Science Laboratory, University College London, Holmbury St.\ Mary, Dorking, Surrey,
RH5 6NT, UK\\
$^2$e2v centre for electronic imaging, The Open University, Walton Hall, Milton Keynes, MK6 7AA, UK\\
$^3$European Space Agency, ESTEC, Postbus 299, 2200 AG Noordwijk, The Netherlands\\
$^4$Leiden Observatory, Leiden University, P.O. Box 9513, 2300 RA, Leiden, The Netherlands\\
$^5$ESA/ESAC, RSSD/{\gaia}, P.O. Box 78, Villanueva de la Ca\~nada (Madrid), Spain\\
$^6$Surrey Satellite Technology Ltd., Sevenoaks, UK\\
$^\dagger${\rm Deceased. This paper is dedicated to Gordon Hopkinson (1952-2010).}}

\begin{document}

\date{Accepted . Received ; in original form \today}

\maketitle

\begin{abstract}

The European Space Agency (ESA) {\gaia} satellite has 106 CCD image sensors which will suffer from  increased charge transfer inefficiency (CTI) as a result of radiation damage. To aid the mitigation at low signal levels, the CCD design includes Supplementary Buried Channels (SBCs, otherwise known as `notches') within each CCD column.  We present the largest published sample of \gaia~CCD SBC Full Well Capacity (FWC) laboratory measurements and simulations based on 13 devices.  We find that \gaia~CCDs manufactured post-2004 have SBCs with FWCs in the upper half of each CCD that are systematically smaller by two orders of magnitude ($\le$50 electrons) compared to those manufactured pre-2004 (thousands of electrons).  \gaia's faint star ($13 \le G \le 20$ mag) astrometric performance predictions by Prod'homme et al. and Holl et al. use pre-2004 SBC FWCs as inputs to their simulations.  However, all the CCDs already integrated onto the satellite for the 2013 launch are post-2004.  SBC FWC measurements are not available for one of our five post-2004 CCDs but the fact it meets \gaia's image location requirements suggests it has SBC FWCs similar to pre-2004.  It is too late to measure the SBC FWCs onboard the satellite and it is not possible to theoretically predict them.  \gaia's faint star astrometric performance predictions depend on knowledge of the onboard SBC FWCs but as these are currently unavailable, it is not known how representative of the whole focal plane the current predictions are.  Therefore, we suggest \gaia's initial in-orbit calibrations should include measurement of the onboard SBC FWCs.  We present a potential method to do this. Faint star astrometric performance predictions based on onboard SBC FWCs at the start of the mission would allow satellite operating conditions or CTI software mitigation to be further optimised to improve the scientific return of \gaia.

\end{abstract}


%
%
\section{Introduction}\label{s:intro}

The European Space Agency (ESA) {\gaia} satellite is a high-precision astrometric, photometric and spectroscopic ESA
cornerstone mission, scheduled for launch in 2013, that will produce the most accurate stereoscopic
map to date of the Milky Way. This will be achieved by measuring  parallaxes, proper motions and astrophysical parameters for one billion stars, one per cent of the estimated stellar
population in our Galaxy \citep{perryman2001} and  radial
velocities for 150 million of these stars. \cite{seabroke2008c} fig. 6 shows that {\gaia}
observations will consist of charge
packets within each CCD pixel ranging from one electron to the pixel Full Well Capacity\footnote{Table~\ref{t:accronym} lists all the acronyms
used throughout the paper along with their definitions.} (FWC) of $190\,000$ electrons. This is due to
{\gaia}'s completeness range ($V \le 20$ mag) and the fixed exposure time for each CCD ($4.4$~s),
due to operating in Time-Delayed Integration (TDI) mode in step with the satellite's spin
rate.\footnote{The variable TDI length is a fundamental part of the {\gaia} system that can be set to keep in step with the satellite spin rate.}  The vast majority of {\gaia} observations will be at the faint end of its magnitude range and
correspondingly the vast majority of the charge packets that make up {\gaia} observations are
expected to range from one to thousands of electrons. {\gaia} will operate in the radiation
environment at L2 for at least five years. 
During the mission, radiation will generate `traps'  in the silicon which will increase charge transfer inefficiency (CTI) i.e. the fraction of the charge signal lost when a charge packet is transferred from pixel to pixel.  The design of the
{\gaia} CCDs includes Supplementary Buried Channels (SBCs) to confine charge packets to a small volume of CCD silicon, thereby reducing the number of traps with which the signal can interact and minimising CTI.

The formally agreed CCD acceptance criterion in the {\gaia} contract between ESA/EADS-Astrium and e2v only stipulated that SBCs should be present in {\gaia} CCDs, not what their FWC should be.  Demonstrator Models (DMs) from the Technology
Demonstration Activities (TDA) phase and Engineering Models (EMs) were tested using the First Pixel Response (FPR) method.  SBCs reveal their presence through a characteristic bump in plots of fractional charge loss as a function of signal size (FPR curves).  The first evidence of this bump in an EM CCD was published in \citet{hopkinson2005}.  We model the same data in detail in this paper.  \citet{hopkinson2006} showed that the characteristic SBC bump was in the majority of CCDs built prior to 2004.  The position of the bumps in the FPR curves indicated that the SBC FWCs are thousands of electrons but SBC FWCs were not systematically measured.
Before manufacturing Flight Models (FMs), e2v changed their photo-lithographic mask set in 2004 but the set was meant to be identical to the one used to manufacture the DMs and EMs.  The mask set was changed to address several important observations made during the TDA phase, which did not include an intentional change of the SBC. SBC FWCs of FMs were not systematically measured either.   

In an independent one-off test, \citet{kohley2009} tested one close-reject post-2004 FM using the pocket pumping and found SBC FWCs $<$40 electrons in the upper half of every CCD column. Note
that this was not the reason for this particular FM not being selected for the actual mission.  \citet{seabroke2010}'s 3D semi-conductor physics model of the {\gaia} pixel shows that an accumulation of nominal photo-lithographic alignment errors furthest from the readout node, thought to be rare by e2v, can reduce SBC FWCs from thousands of electrons to zero.  We present the largest published sample of \gaia~CCD SBC Full Well Capacity (FWC) laboratory measurements and simulations based on 13 devices to investigate whether this post-2004 CCD is a rare example of the compounding photo-lithographic alignment errors or whether it is typical of post-2004 CCDs, which would point to systematic differences between the SBC FWCs in pre- and post-2004 CCDs.  The change in the mask set can only be considered as circumstantial evidence to explain any systematic changes between pre- and post-2004 CCDs because it would be very difficult to definitively prove the mask set was responsible due to a plethora of other factors that go into manufacturing CCDs with complex pixel architectures.  



Reduced SBC FWCs would
increase the effects of CTI on {\gaia} observations. This introduces a systematic bias
in the measurements (e.g.\ image location on the CCD, radial velocities etc). As described in detail
in \cite{prodhomme2011b} the {\gaia} data processing takes into account CTI effects via the forward
modelling of the image distortion. This approach allows for the unbiased estimation of the stellar
image parameters. However, the nature of {\gaia}'s observations (TDI and windowed - only a small
region around the image or spectrum is read out from the CCD and telemetered to the ground) means
that CTI also induces signal loss. This signal loss is expected to degrade the
predicted faint star ($13 \le G \le 20$ mag) astrometric performance of the mission. For CCDs with pre-2004 SBC FWCs, 
\cite{prodhomme2011b} predict a faint star astrometric performance degradation (over the CTI-free case) of about 10 per cent. With
reduced SBC FWCs this performance loss will be higher.

{\gaia}'s CCDs have already been integrated on to the satellite and so it is too late to change CCDs which are
selected to fly on {\gaia}.  Nevertheless and in view of the CTI calibration, the enormous amounts of data that {\gaia} will produce in its 5-year
mission means it is important to establish the
potential impact on the mission of reduced SBC FWCs. This paper addresses the following issues:

\begin{enumerate}
 \item In the absence of testing FM CCD SBC FWCs as a criterion for selecting
    which CCDs should fly on the {\gaia} satellite, is it possible to predict how many will have
    reduced FWCs?
  \item Are the SBC FWCs in {\gaia} CCDs used to predict the faint star astrometric performances
    and against which mitigation models are being developed and tested representative of the CCDs that will fly on \gaia?
  \item What is the impact of reduced SBC FWCs on the {\gaia} image location accuracy?
\end{enumerate}

The paper is organised as follows: the {\gaia} CCD and SBC is introduced in detail in Section~\ref{s:sbc}.  Section~\ref{s:expt} re-analyses the \cite{hopkinson2006} data set of seven CCDs and models the FPR data from two CCDs: a typical case and a rare compounding photo-lithographic alignment error case.  We conduct our own pocket pumping measurements and analysis of two {\gaia} CCD test structures in Section \ref{s:pocket}.  Section~\ref{s:discuss:rc} presents a re-intepretation of data obtained from irradiated devices tested by EADS-Astrium, {\gaia}'s main industrial
partner.  We present and discuss the answers to the questions raised above in Section \ref{s:discuss}.  The conclusion in Section \ref{s:conc} suggests some future work in Section \ref{s:recom}, which is detailed in the Appendix.

\section{The Gaia CCD architecture} \label{s:sbc}

\begin{figure}
\centering
  \includegraphics[width=0.85\columnwidth]{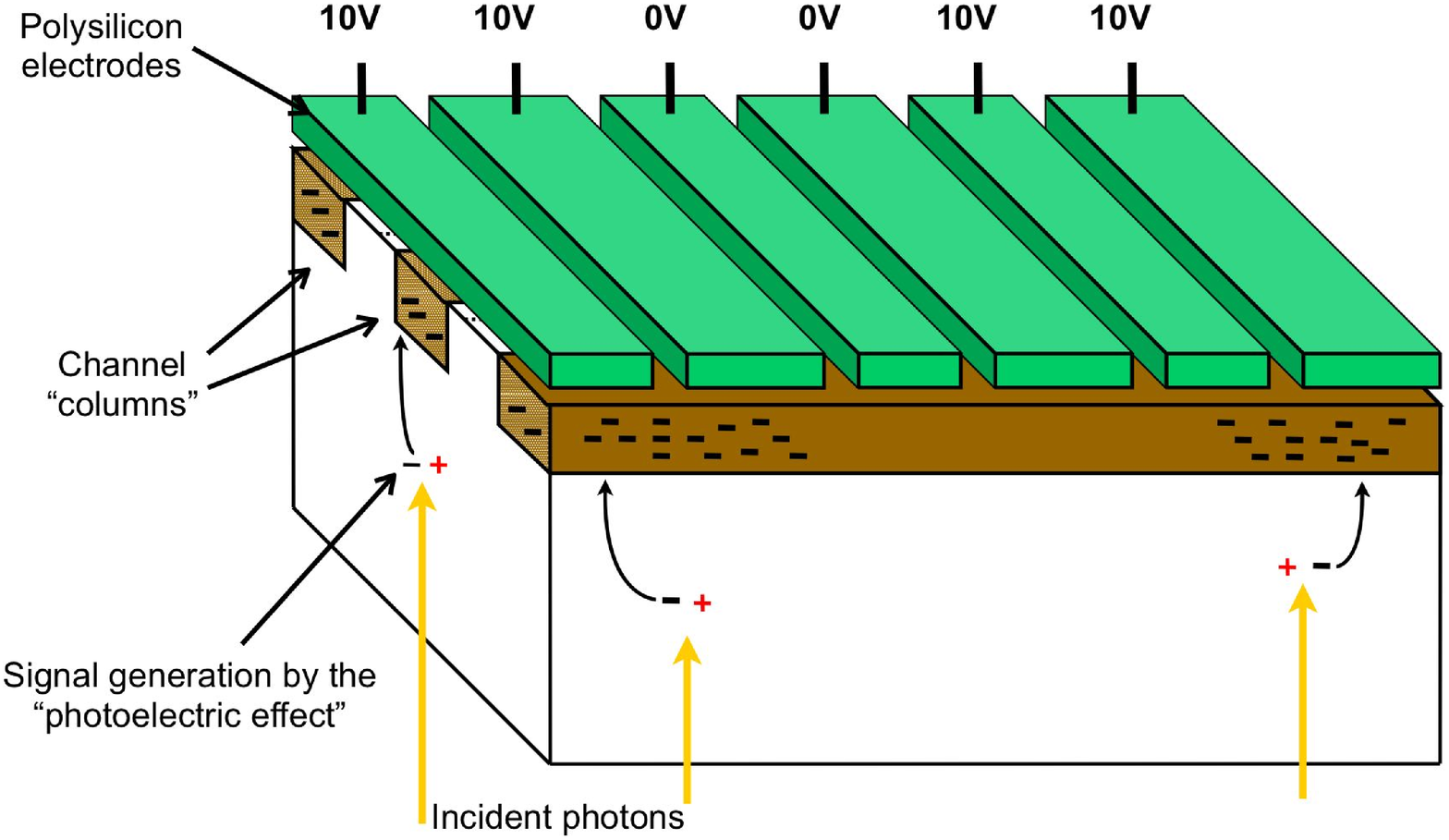}
  \caption{Front-side schematic of a {\gaia} CCD showing the four-phase electrode structure.  {\gaia} CCDs are back illuminated i.e. incident photons do not pass through the electrodes.  Two consecutive electrodes are biased `high' (10 V), to which photoelectrically generated signal electrons are attracted.  Charge packets are separated in the charge transfer direction by biased `low' (0 V) electrodes.  The cube below the electrodes is a block of silicon and the brown regions within the silicon are implanted doping, which form channel columns.  Charge packets are separated perpendicular to the charge transfer direction by column isolation regions of undoped silicon.}
  \label{f:bil}
  \includegraphics[width=0.85\columnwidth]{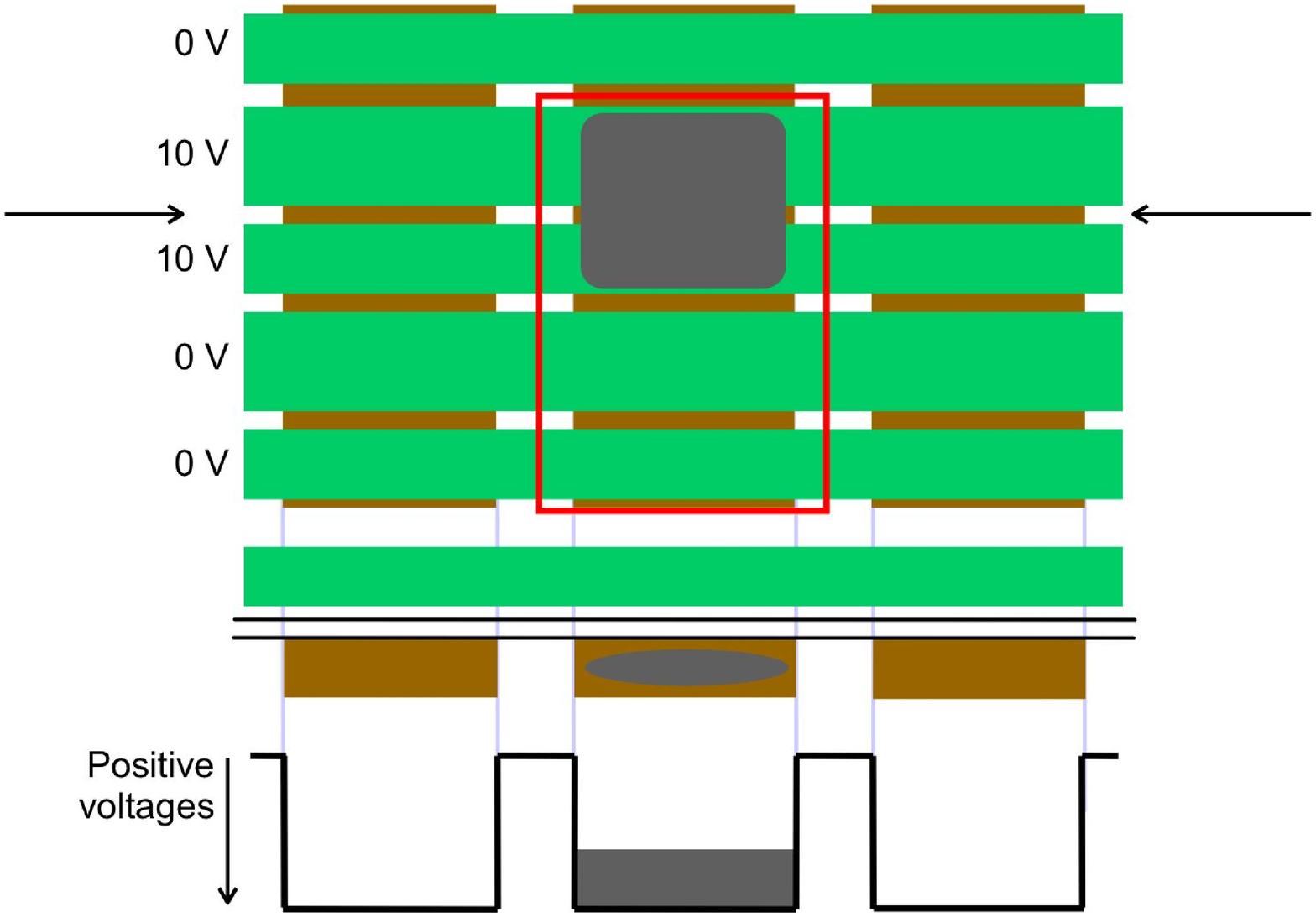}
  \caption{{\it Top}: Front-side schematic of a CCD showing BCs in brown vertical rectangles, electrodes in green horizontal rectangles, a charge packet in the rounded grey region with a pixel illustrated with a red rectangular outline.  {\it Bottom}: Cross-section through the arrows in the above schematic showing a single electrode above three BCs with the charge packet in the middle BC.  A simplified potential distribution is below, showing that the charge packet sits within the potential maximum of a BC.}
  \label{f:bc}
  \includegraphics[width=0.85\columnwidth]{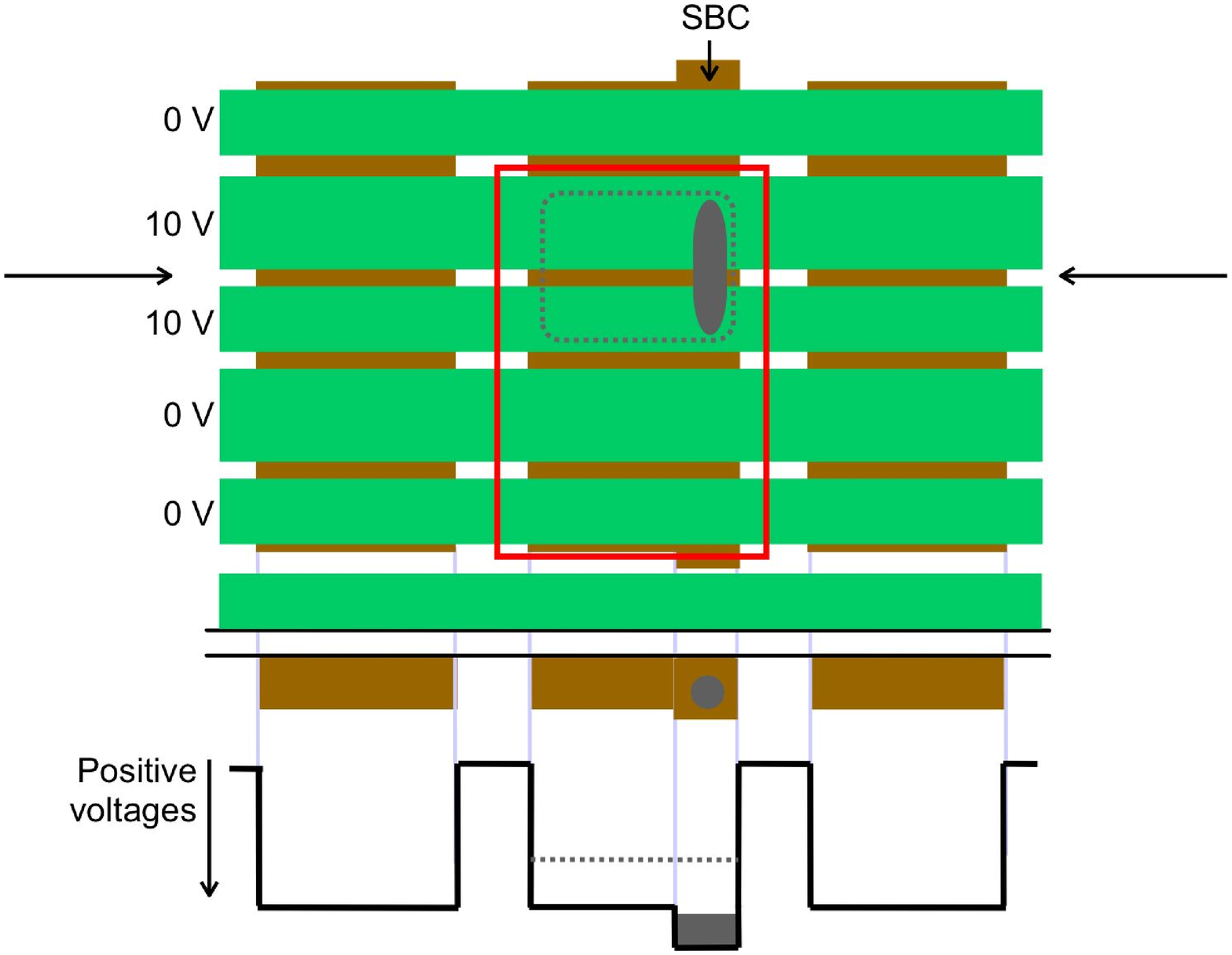}
  \caption{Same as Fig. \ref{f:sbc} but with SBC doping added to the central BC.  The charge packet illustrated is smaller than the SBC FWC and so sits in the SBC volume, rather than the BC volume({\it top}) and sits in the higher SBC potential maximum, rather than the lower BC potential maximum ({\it bottom}).}
  \label{f:sbc}
\end{figure}


Fig.~\ref{f:bil} illustrates the general principles of the {\gaia} CCD.  Buried Channels (BCs) and SBCs are introduced in Figs.~\ref{f:bc} and \ref{f:sbc} respectively.  Once the SBC FWC is reached, any further charge spreads out to cover the whole BC volume, as shown by the dotted lines in Fig. \ref{f:sbc}.


\begin{figure}
  \includegraphics[width=\columnwidth]{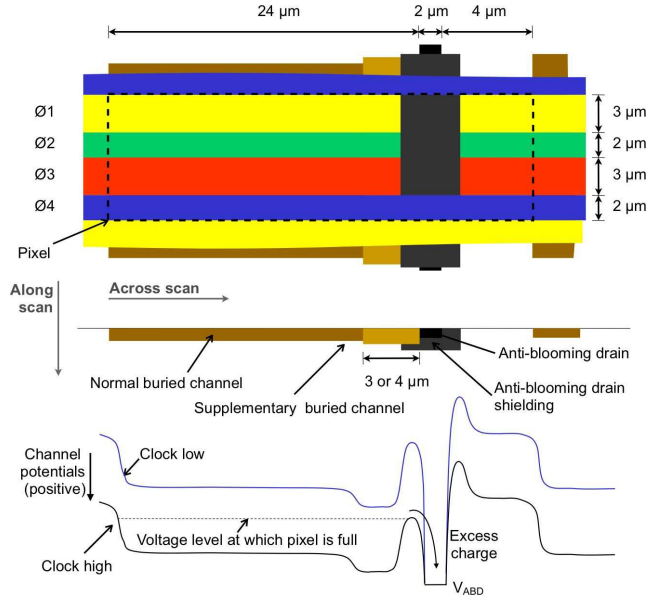}
  \caption{Detailed schematic of the {\gaia} CCD pixel architecture. {\it Top}: Front-side schematic of a {\gaia} CCD pixel showing the four electrodes
  ($\phi$) with pixel features labelled. {\it Middle}: Schematic of the vertical cross-section
  through the top schematic in the across-scan direction. {\it Bottom}: Channel potential profile in
  the same direction, resulting from different voltages being applied to the electrode (clock low $=
  0$ V, clock high $= 10$ V).}
  \label{f:pixel}
\end{figure}


\begin{figure}
  \includegraphics[width=\columnwidth]{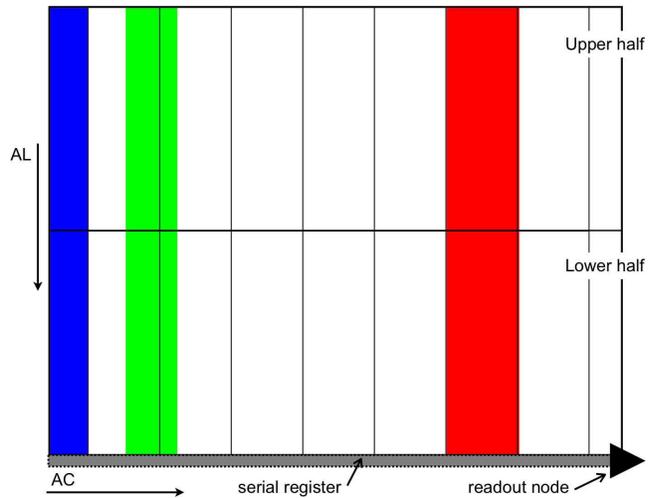}
  \caption{Schematic of the {\gaia} CCD image area: CCD columns (TDI line ALong scan, AL) are
  vertical; ACross scan (AC) is horizontal; the readout node is at the bottom right corner.
  Each rectangle is a stitch block. All the internal lines in the schematic are stitch boundaries.
  The horizontal stitch boundary is at TDI line 2160 from the serial register (bottom box) and at TDI line 2340 from the start of the CCD (top line in the schematic).  The three colour-coded areas
  are referred to in the text: (i) the red (rightmost) 
  area covers columns 359-608, corresponding to two stitch blocks in the AL direction, hereafter
  known as AL stitch block couples (which corresponds
  to the AL stitch block couple identified in 03483-05-02 as typical for the pre-2004 sample, see Fig. \ref{f:fitexamples} right); (ii) the blue (leftmost) area covers columns 1859-1966 (which corresponds
  to the AL stitch block couple identified as a candidate for reduced SBC FWCs in 03153-20-01, see Fig. \ref{f:fitexamples} left);  (iii) the green (middle) area covers columns 1550-1755 (which corresponds to the area over
  which 05256-17-02 was irradiated, see Fig. \ref{f:rc_bias}).}
  \label{f:blocks}
\end{figure}

The CCD91-72 was designed and manufactured by e2v especially for {\gaia}.  {\gaia} has one of most complex pixel architectures built for astronomy (see Fig. \ref{f:pixel}).  It includes the standard BC, the relatively rare SBC and, unusually, an Anti-Blooming Drain (ABD).  {\gaia} is the only astronomical detector with an ABD.  It prevents charge bleeding down columns from bright observations, allowing simultaneous faint observations.  The ABD also removes excess charge from just upstream of TDI gates.  These  gates, located at different positions within each CCD, block charge packets of very bright observations that would otherwise saturate the pixels.  This allows their integration to begin just downstream of the TDI gate at shorter distances from the readout register.  

The doping that defines each pixel feature is implanted into the CCD silicon using a photo-lithographic mask. Each feature
(e.g. BC or SBC) has its own mask. Due to practical manufacturing constraints, the masks are smaller
than the image area of large format devices like the {\gaia} CCD (4.5 cm $\times$ 5.9 cm). 
The mask area dictates the area of the CCD over which the photoresist layer can be patterned at any one time. 
This smaller sub-array is called a stitch block \citep[][see
Fig.~\ref{f:blocks}]{bruijne2008}. Large format devices like the {\gaia} CCD are fabricated using a
photo-lithographic `step and repeat' process that patterns the photoresist for a particular pixel feature simultaneously everywhere in a
stitch block, one stitch block at a time.  After the entire wafer has been patterned a single ion implantation process is performed.

Both {\gaia} CCD's upper and lower halves are
photo-lithographically stitched from 7 repeated stitch blocks and 2 end-termination stitch blocks. A
{\gaia} CCD is thus composed of 18 different stitch blocks.  Each termination stitch block
contains 108 columns while each repeated stitch block section contains 250 columns: $(2\times108) +
(7\times250) = 1966$ columns in the ACross scan (AC) direction and 4500 TDI lines in the ALong scan
(AL) direction. The AL direction refers to the scan direction of stars along {\gaia} CCD columns that is induced by the continuous spinning of the spacecraft around its own axis. 

In the CCD upper half (cf.\ Fig.~\ref{f:blocks}) the nominal SBC doping width is 3 $\mu$m, whereas in
the CCD lower half the nominal SBC doping width is 4 $\mu$m \citep{burt2003}. This is because every
CCD column and so every SBC crosses the horizontal stitch boundary. The SBC doping width is
increased immediately after the stitch boundary to minimise the possibility of the SBC narrowing at
the boundary as it is known that boundaries can cause the formation of potential pockets, which
act like radiation traps, capturing and then releasing electrons and thus increasing CTI. This
difference between the nominal SBC doping widths in the upper and lower halves of each {\gaia} CCD
translates to different SBC FWCs in either half. 

\section{First Pixel Response measurements}\label{s:expt}


%
%

\subsection{Technique} \label{s:expt:fpr}

The FPR measurement consists of the analysis of the charge loss (induced by trapping) that occurs in
the first pixels of a well-characterized signal after its transfer across the full CCD image area.
By well-characterized signal, we mean a signal for which the shape and the number of charges can be inferred independently from the CTI effects such as for a CI.

In the first pixel row only, the {\gaia} CCD pixels comprise a diode to generate artificial
charges and a gate to control the number of artificial charges injected i.e. effectively transferred
across the CCD (cf. appendix for more details about charge injection techniques). During the {\gaia} mission, CIs will be performed
periodically (every $\sim$1 s) by blocks of 4 to 20 lines to fill a large fraction of the traps prior to the stellar transits and thus mitigate the CTI effects.

\begin{figure}
\begin{center}
 \includegraphics[angle=-90,width=\columnwidth]{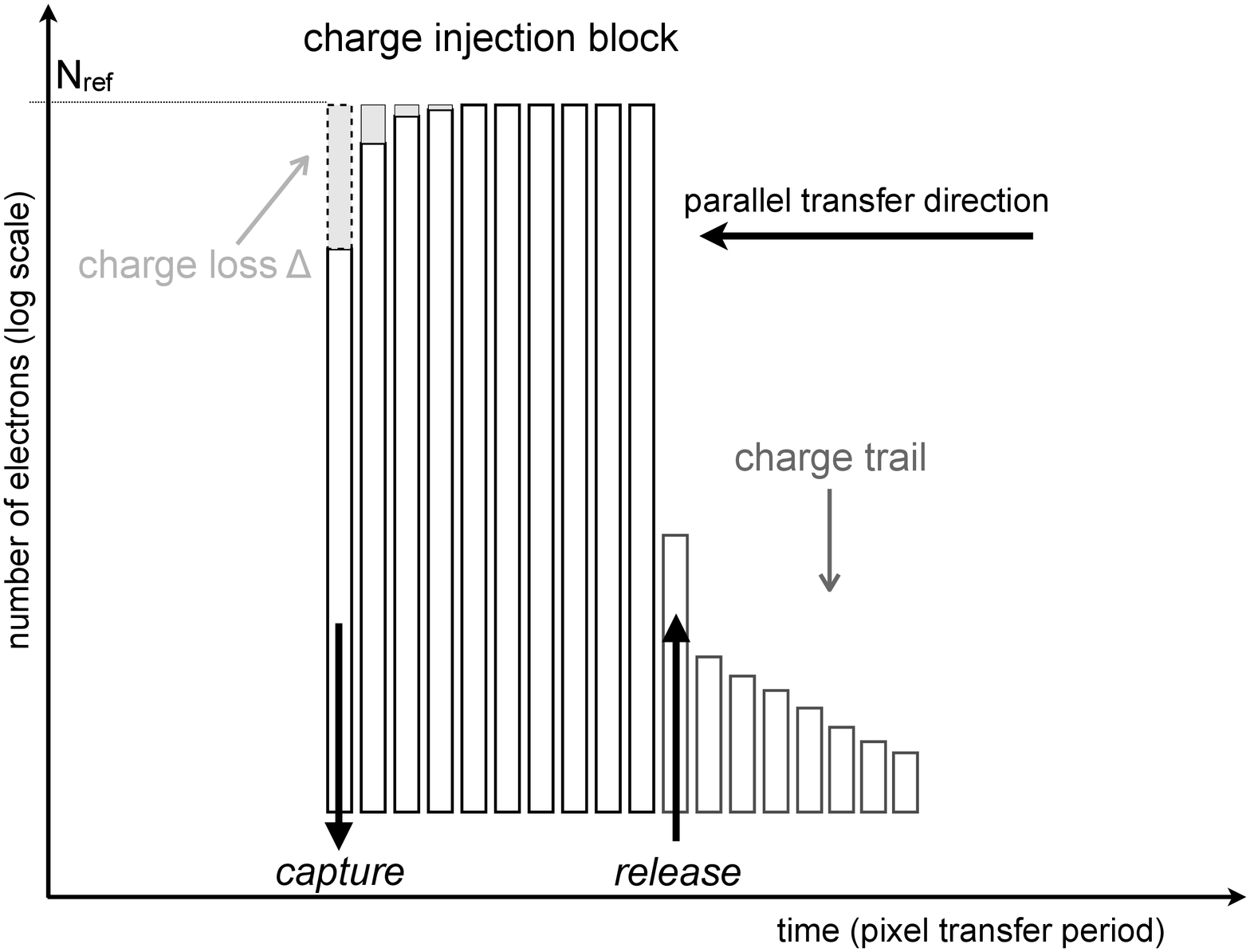}
 \end{center}
 \caption{Schematic presenting the principle of a FPR measurement. The charge loss $\Delta$ occurring in the first pixel(s) of a CI is computed using the measured CI reference level $N_{ref}$.}
 \label{f:fpr_explained}
\end{figure}

As depicted in Fig. \ref{f:fpr_explained}, in a CI block of several tens of lines, the first line undergoes the most damage by encountering a certain fraction of the total amount of
active traps (i.e.\ the empty traps) present in the signal confinement volume. As we shall see this
fraction depends on the CI level (the number of electrons per pixel in the CI block) and the
electron density distribution within a pixel. Depending on the capture cross-section of the trap
species present and the clocking rate, the first CI line may not fill all the encountered traps.  This is also because the first CI line is losing signal along each CCD column as charge is captured. 
In this case the second CI line will also experience charge loss and subsequently for the other
lines of the CI block. After a certain number of lines, no significant trapping can be measured.
Using the last lines of a CI block one can measure a reference CI level, $N_\mathrm{ref}$, and thus
compute the charge loss $\Delta$ (i.e.\ the total number of trapped charges), and the fractional
charge loss $\delta$:
\begin{gather}
  \Delta = \sum_{i=0}^{S-1} N_\mathrm{ref} -  N_i\,, \\[3pt]
  \delta = \frac{\Delta}{S\,N_\mathrm{ref}}\,,
  \label{eq:fcl}
\end{gather}
where $N_i$ is the number of \electron pixel$^{-1}$ in the $i^\mathrm{th}$ line of the CI block.  The traps
filled by the CI with a release time constant shorter than the CI block duration will release
electrons within the CI block and can thus bias the measurement of the CI reference level.  To
minimize this source of uncertainty only the first line(s) of the CI block should be considered in the
charge loss measurement. The charge loss measured in the first lines of a CI block depends on the
state of the traps at the time they are encountered by the CI signal. The trap state is set by the
CCD illumination history. During a FPR experiment, it is thus important to have a good control (or
at least a good knowledge) of the CCD illumination history. In the FPR experiments carried out by
\cite{hopkinson2005} to characterize the effect of the SBC, two CI blocks were performed with 100
lines of delay between them. The first CI block was performed to reset the illumination history, and
only the second block was used to measure the charge loss. The level of the first block was fixed to
$30\,000$ \electron, while the level of the second block was varied. In this way, one can study the
charge loss variation as a function of signal level which enables the characterization of the SBC within a CCD pixel column. Only the first pixel was taken into account in the fractional charge loss measurement.

%
%

\subsection{Selection of the data}\label{s:expt:selection}

\begin{figure}
\begin{center}
 \includegraphics[angle=-90,width=\columnwidth]{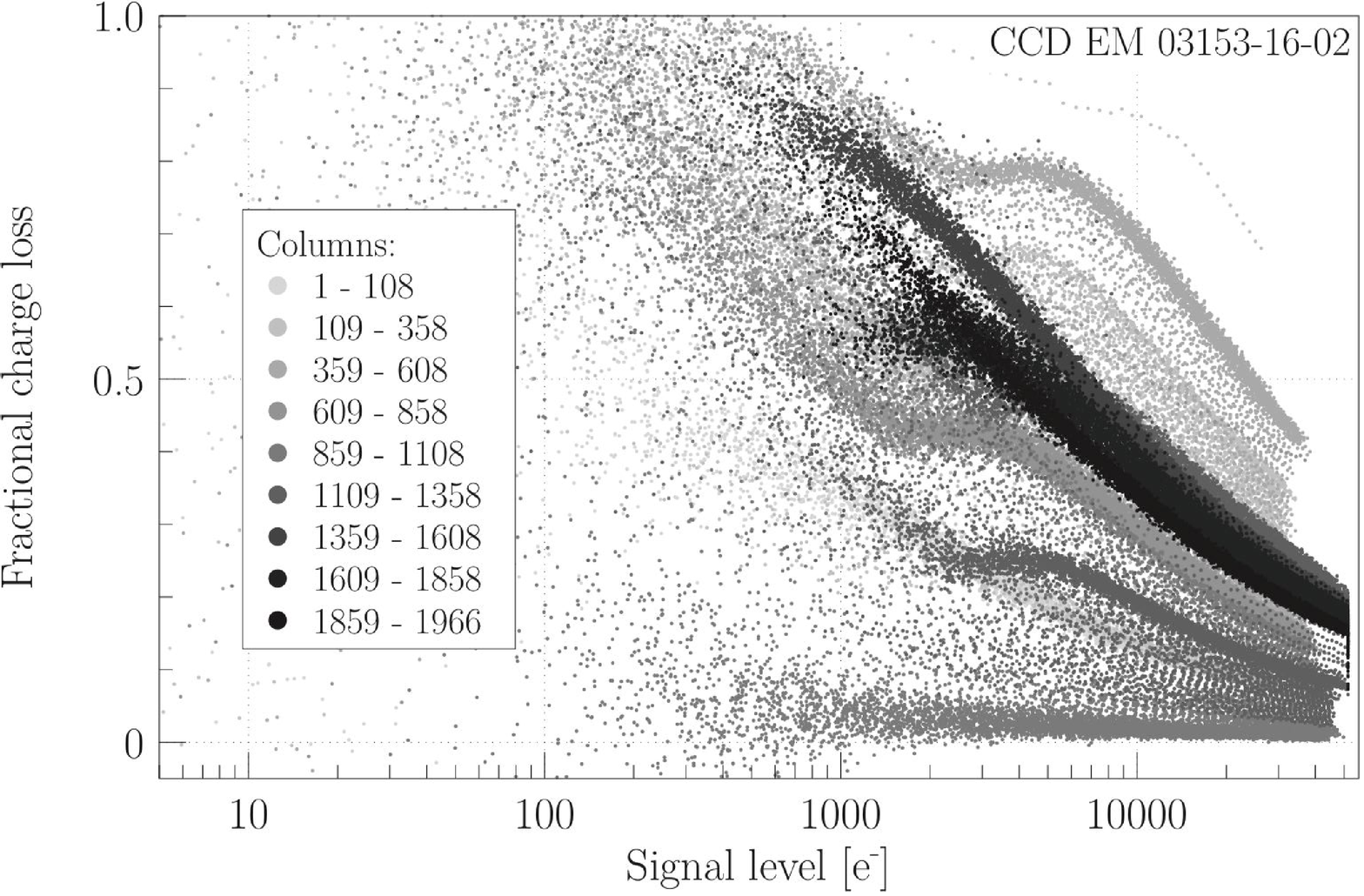}
 \includegraphics[angle=-90,width=\columnwidth]{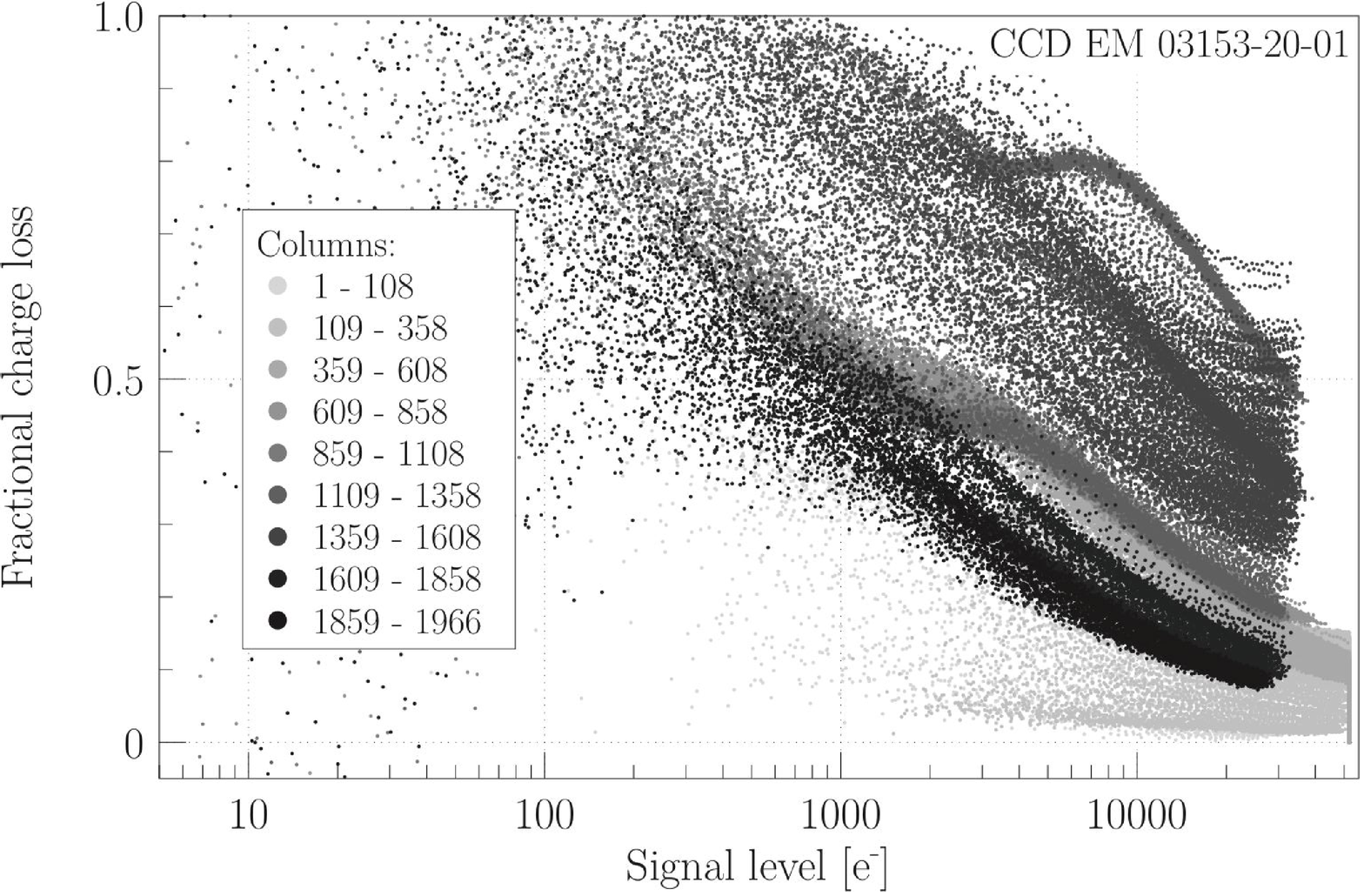}
 \includegraphics[angle=-90,width=\columnwidth]{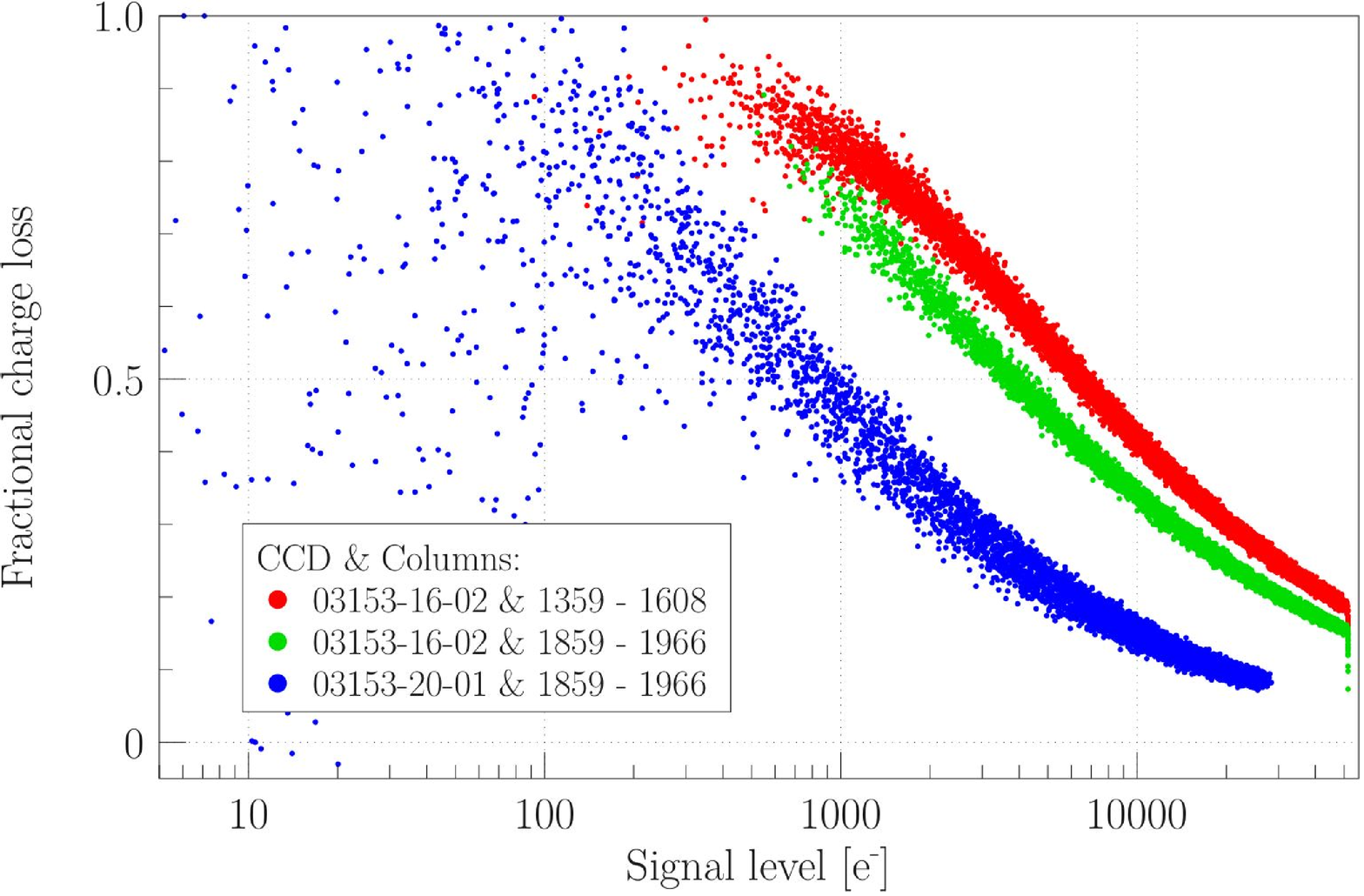} 
 \end{center}
 \caption{{\it Top and middle}: FPR plots of the two CCDs out of the seven including AL stitch block
 couples that are candidates for reduced SBC FWCs. {\it Bottom}: FPR plots of the three AL stitch block
 couples that are candidates for reduced SBC FWCs.}
 \label{f:3egs}
\end{figure}

The first set of CCD radiation
test data obtained by an industrial partner in the {\gaia} project was by Sira. These electro-optics tests were carried out on seven {\gaia} Astrometric Field (AF) DM and EM CCDs (see Table~\ref{t:sira_CCDs}) and are described in
\cite{hopkinson2005} and \cite{hopkinson2006}.   \cite{hopkinson2006} figs.~8-2-2 and 8-2-3  show FPR data from different
columns from different AL stitch block couples.  The top two plots in Fig. \ref{f:3egs} group these data into AL stitch block couples for two of these CCDs.  The majority of the AL stitch block couples exhibit a
characteristic bump in their fractional charge loss curves.  This bump is due to SBCs.  

At the leftmost (smaller signal level) inflexion point in these fractional charge loss curves, the number of electrons in
the SBC start to collapse the SBC potential so some of these electrons spill out of the SBC
potential into the BC potential. No longer protected in the smaller volume of the SBC, these few
electrons in the BC meet more traps, causing the fractional charge loss increase. At the rightmost (larger signal level)
inflexion point, the signal level has completely collapsed the SBC potential into the BC potential
so here the SBC potential well no longer exists and all the electrons sit in the BC potential.  The presence of SBCs causes the fractional charge loss curve
to shift left to smaller signal levels so that at a given signal level the fractional charge loss in
the SBCs is less than it would be in the BCs.

We visually examined the fractional charge loss curve of every AL stitch block couple in each of the seven CCDs and found three examples from two CCDs that do
not exhibit the characteristic SBC bump (see bottom plot of Fig. \ref{f:3egs}).  It was decided to model the AL stitch block couple that had FPR data extending to
the smallest signal levels (EM 03153-20-01 columns 1859--1966, see Fig.~\ref{f:blocks} blue area) to provide more modelling constraints.  The other two examples show hints of turning over into the characteristic SBC bump where their data stops.  In the next section, we model this data to test our hypothesis that a fractional charge loss curve without a bump indicates a reduced SBC FWC. 



\begin{table}
  \centering
  \caption{Summary of the irradiated AF CCDs tested by Sira. The first five digits of the CCD serial
  number form the batch number.  The digits in the batch number refer to when the front-side processing occurred (explained in Section \ref{s:discuss:stats}).  The first two digits in the batch number are the year e.g. 03 refers to 2003.  The second two digits in the batch number are the week of the year e.g. 15 out of 52.  The fifth digit in the batch number is the number of batches in that week e.g. 3rd batch of week 15 in 2003.  The middle set of two digits in the serial number gives the wafer
  number. The other set of two digits refers to the position of the CCD within the wafer. 
  }
  \begin{tabular}{|c|l|}
	\hline
	Serial Number 	& Model (Extra information)\\
	\hline
	03153-05-02                          & DM (front illuminated)\\
	03153-07-01                          & DM (front illuminated)\\
	03153-16-02                          & DM\\
	03153-20-01                          & DM\\
	03442-11-01                          & EM\\
	03483-05-02                          & EM\\
	03483-06-02                          & EM\\
	\hline
	\end{tabular}
\label{t:sira_CCDs}
\end{table}


\subsection{Generation of the synthetic data} \label{s:expt:sim}
\subsubsection{Simulation setup}\label{s:expt:sim:distribmodel}
In order to reproduce the FPR measurements, we simulate the transfer of a CI block over the image
area of a virtual irradiated CCD using the detailed Monte-Carlo model of charge transfer presented
by \cite{prodhomme2011}\footnote{This model can be used as part of the Java package {\cemga}
available online: \url{http://gaia.strw.leidenuniv.nl/cemga}}. 
This model was
verified against experimental data acquired with {\gaia} irradiated CCDs. In particular, by using a
flexible and analytical representation of the electron density distribution, \cite{prodhomme2011}
were able to accurately reproduce the measurement of fractional charge loss as a function of signal
level in the presence of SBCs (in both CCD halves). 
As already mentioned the SBC confines the electron packet in depth ($z$) and in the AC
direction ($y$), but the packet remains spread under two electrodes in the parallel transfer
direction ($x$) independently of the signal regime. Thus, in the model, the SBC and BC signal
confinement volumes have different dimensions along $y$ and $z$ ($y_\mathrm{max}$,
$z_\mathrm{max}$, $y_{\mathrm{SBC},\mathrm{max}}$, and $z_{\mathrm{SBC},\mathrm{max}}$), but the same size along $x$ ($x_\mathrm{max}$). Accordingly the centre and
the standard widths of the density distribution along $y$ and $z$ are different for the BC and
SBC regimes. For each regime, the distribution standard width in one direction is set to a
fraction $\eta$ of the signal confinement volume dimension in that direction, e.g.: $\sigma_{y} = \eta \times y_\mathrm{max}\,$.
In this way the ratio between the dimensions of the signal confinement box is preserved in the
electron density distribution. A larger $\eta$ means a larger spread in electron distribution.
$\eta$ is thus later referred to as the electron density distribution spread factor.

Like in \cite{prodhomme2011}, the BC signal confinement volume and the standard widths of the electron
density distribution remain the same for the two CCD halves.  \cite{prodhomme2011} modelled a single set of SBC parameters in both CCD halves.  However, as explained in Section \ref{s:sbc}, the SBCs have different doping widths in each CCD half and so should have different values of $\eta$ and FWCs ($S_{\mathrm{SBC}}$) in each CCD half.  This more realistic modelling is the main difference between the  simulation in \cite{prodhomme2011} and the one in this paper.  Therefore, in this paper $\eta_1, \eta_2, S_{\mathrm{SBC}_1}$, and $S_{\mathrm{SBC}_2}$ are free parameters in the model, where 1 is the CCD upper half (furthest from the readout register) and 2 is the lower half.  

\begin{figure}
    \includegraphics[width=0.49\textwidth]{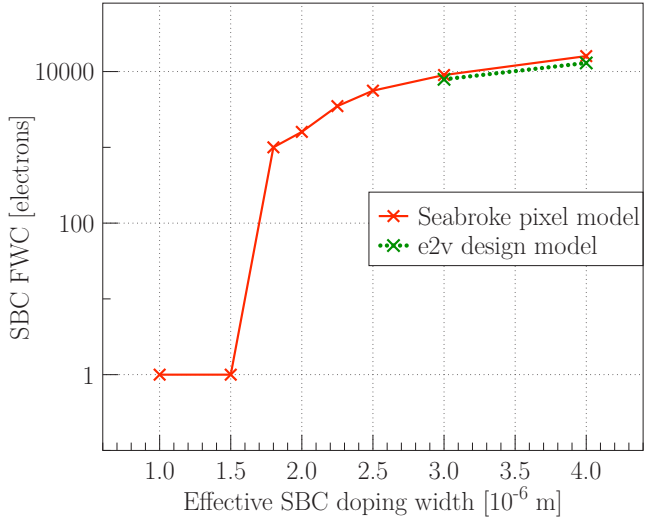}
  \caption{\citet{seabroke2010} pixel model and e2v design model predictions of SBC FWC as a function of effective SBC doping width (same data plotted as in \citealt{seabroke2010} fig. 5).}
  \label{f:weff}
\end{figure}

The nominal SBC doping width is 3 $\mu$m in the upper half and 4 $\mu$m in the lower half.  If the SBCs are undisturbed in the pixel, one would expect $y_{\mathrm{SBC},\mathrm{max}}$ to be similar widths.  However, as shown in Section \ref{s:sbc}, the SBCs are not undisturbed: they are adjacent to ABDs.   A 3D semi-conductor physics
model of the {\gaia} pixel \citep{seabroke2010} finds that ABDs reduce the nominal SBC doping widths to effective SBC doping widths ($w_{\textrm{eff}}$) and thus also reduce $y_{\mathrm{SBC},\mathrm{max}}$.   Fig. \ref{f:weff} shows that $(w_{\textrm{eff,1}},w_{\textrm{eff,2}}) = (3,4)$~$\mu$m yield simulated SBC FWCs of $\sim$10~000 electrons.  The bottom plot of Fig. \ref{f:3egs} shows that our candidate reduced SBC FWC AL stitch block couple does not exhibit a SBC bump near 10~000 electrons.  Therefore, a more realistic constraint on $y_{\mathrm{SBC},\mathrm{max}}$ for this particular simulation is 2 $\mu$m.  This corresponds to a maximum SBC FWC of several 1000 electrons. 

The transition between the two signal regimes occurs at $S_\mathrm{SBC}$. It is important to note that
although in reality the model parameter $S_\mathrm{SBC}$ would be equated to the real SBC FWC, in the
model used, due to the arbitrary nature of the chosen description for the transition between the BC and
SBC regimes, $S_\mathrm{SBC}$ gives only an indication about the real value of the SBC FWC (see \cite{prodhomme2011} for more details).

During the experiments carried out by Hopkinson, two CI blocks were transferred. The first CI block
was used to reset the illumination history by filling a large fraction of the empty traps. The
second block was used to perform the FPR measurement and its level varied. For most of the tested
levels, the level of the first block ($\sim$30$\,$000~\electron~pixel$^{-1}$) was higher than the
level of the second. As a consequence the population of traps probed by the second CI block was the
same as for the first. However from this population the only traps capable of capturing electrons
from the second block, are by definition the empty traps: the ones that were able to release their
electron before the crossing of the second block. In our simulated experiment we can thus simulate
the transfer of a single CI block (of 20~lines) crossing the image area of a CCD containing only
empty traps and perform the FPR measurement on this very same CI block. As for this particular
experiment we are interested in the electron capture only, we used a unique trap species with a
common capture cross section ($\trapsigma$~=~5$\times$10$^{-16}$~cm$^2$) and a long release time
constant compared to the duration of the CI (20~$\times$~0.9892~ms) such that the charge loss and
the reference level measurements are not biased by a significant release of electrons.

The simulated
CCD image area consists in a single column of 4500~pixels. We did not simulate the transfer across
the serial register, as one can to first order ignore the effects of the serial CTI.
Once the CI block is transferred through the two CCD halves, the FPR measurement is performed in the
same way as for the experimental test (Section \ref{s:expt:fpr}). For each  set of simulation
parameters, the CI level is varied from 50 to $40\,000$~\electron (with $100$, $200$, $500$, $1000$, $2000$,
$5000$, $10\,000$, $20\,000$ \electron as intermediate values).

%
%

\subsubsection{Model to data comparison}\label{s:expt:sim:comparison}
In our simulations, there are six free parameters:
\begin{enumerate}
  \item  $\eta_{\mathrm{BC}}$, $\eta_1$ and $\eta_2$ the electron density distribution spread
    factors, that set the standard widths of the electron distribution for each signal regime and
    CCD half according to the signal confinement volume dimensions,
  \item $S_{\mathrm{SBC}_1}$ and $S_{\mathrm{SBC}_2}$ the signal level at which the transitions
    between the SBC and BC regimes is performed,
  \item $N_t$ the number of traps per pixel.
\end{enumerate}
Table \ref{t:param} details all the simulation parameters and indicates the values we used for the
fixed parameters as well as the allowed intervals for the free parameters.

We are not interested in the exact reproduction of the data but rather in the study of the
parameters of the electron density distribution model that lead to a reasonable agreement between
the simulated and the experimental fractional charge loss measurements over the studied range of CI
signal levels.  We thus generate random sets of parameters in order to probe homogeneously the
entire parameter space. Each parameter set results in a set of data points that sample the simulated
fractional charge loss curve. These data points are then compared to the experimental measurements.
To proceed to this comparison, we have first generated an analytical representation of the
experimental data by the mean of a fit using spline functions. We use the analytical fit values $F$
and the simulation results $\phi$ at each particular signal level $n$ to compute the $\chi^2$, that
constitutes our comparison criterion:

\begin{equation}
  \chi^{2} = \sum_{n=0}^{N-1} {\frac {\left( \phi\left(n \right) - F\left(n \right)
  \right)^{2}}{\sigma^2}}\,,
\end{equation}
where $N$ is the total number of simulated CI levels and $\sigma$ the noise. We use the formal errors
associated to the analytical fit to the experimental data as $\sigma$ values and thus assume that
the formal errors encompass the experimental noise and the readout noise.

\begin{table}
  \resizebox{0.49\textwidth}{!}{
  \centering 
    \begin{tabular}{|l|l|l|} 

    \hline
    Parameter 	& Description 						& Fixed value or $[$interval$]$ 		\\ \hline
    T 			& temperature						& 163 K 	\\
    P$_{\mathrm{TDI}}$& TDI period 				& 0.9828 ms\\
    $N_t$		& number of traps per pixel		&  $[0$--$6]$\\
    \trapsigma	& capture cross section				& 5$\times10^{-20}$ m$^2$\\
    $\tau_r$		& release time constant				& 9 s\\
    \hline

    & \textbf{BC regime}		 			& 	\\
    \hline
    $\eta_{\mathrm{BC}}$& distribution spread factor 		&$[0.05$--$1]$\\ 
    $S_\mathrm{BC}$ 				& BC FWC 			& $190\,000$ $\electron$ \\
    &\multicolumn{2}{|l|}{\it BC signal confinement volume size:} \\
    $x_\mathrm{max}$		& in the AL direction		& 11 $\mu$m \\
    $y_\mathrm{max}$		& in the AC direction		& 24 $\mu$m  \\
    $z_\mathrm{max}$		& in depth								& $0.75$ $\mu$m \\
    \hline

    & \textbf{SBC regime}		 		& \\
    \hline
    &\multicolumn{2}{|l|}{\it distribution spread factor:} \\
    $\eta_1$  & in the CCD upper half 				& $[0.05$--$1]$ \\
    $\eta_2$  & in the CCD lower half 				& $[0.05$--$1]$ \\
    &\multicolumn{2}{|l|}{\it SBC to BC regime transition signal:} \\
    $S_{\mathrm{SBC}_1}$  & SBC FWC in the CCD upper half 				& $[1$--$8000]$ \electron \\
    $S_{\mathrm{SBC}_2}$  & SBC FWC in the CCD lower half 			& $[1$--$8000]$ \electron \\
    &\multicolumn{2}{|l|}{\it SBC signal confinement volume size:} \\
    $y_{\mathrm{SBC},\mathrm{max}}$		& in the AC direction		& 2 $\mu$m  \\
    $z_{\mathrm{SBC},\mathrm{max}}$		& in depth								& $0.1$ $\mu$m \\
    \hline
  \end{tabular}
  }
   \caption{Simulation parameters.   Although Fig. \ref{f:pixel} shows each pixel is $3 + 2 + 3 + 2 = 10$ $\mu$m in the AL direction, there is actually 0.25 $\mu$m between each electrode and so $x_{\mathrm{max}} = 3 + 0.25 + 2 + 0.25 + 3 + 0.25 + 2 + 0.25 = 11$ $\mu$m.  \citet{seabroke2010} figs. 3 and 4 show that the depth of the SBC is slightly larger than the BC.}
  \label{t:param}	
\end{table}

%
%

\subsection{Comparison results} \label{s:expt:results}
\begin{figure*}
  \centering
  \includegraphics[width=0.49\textwidth]{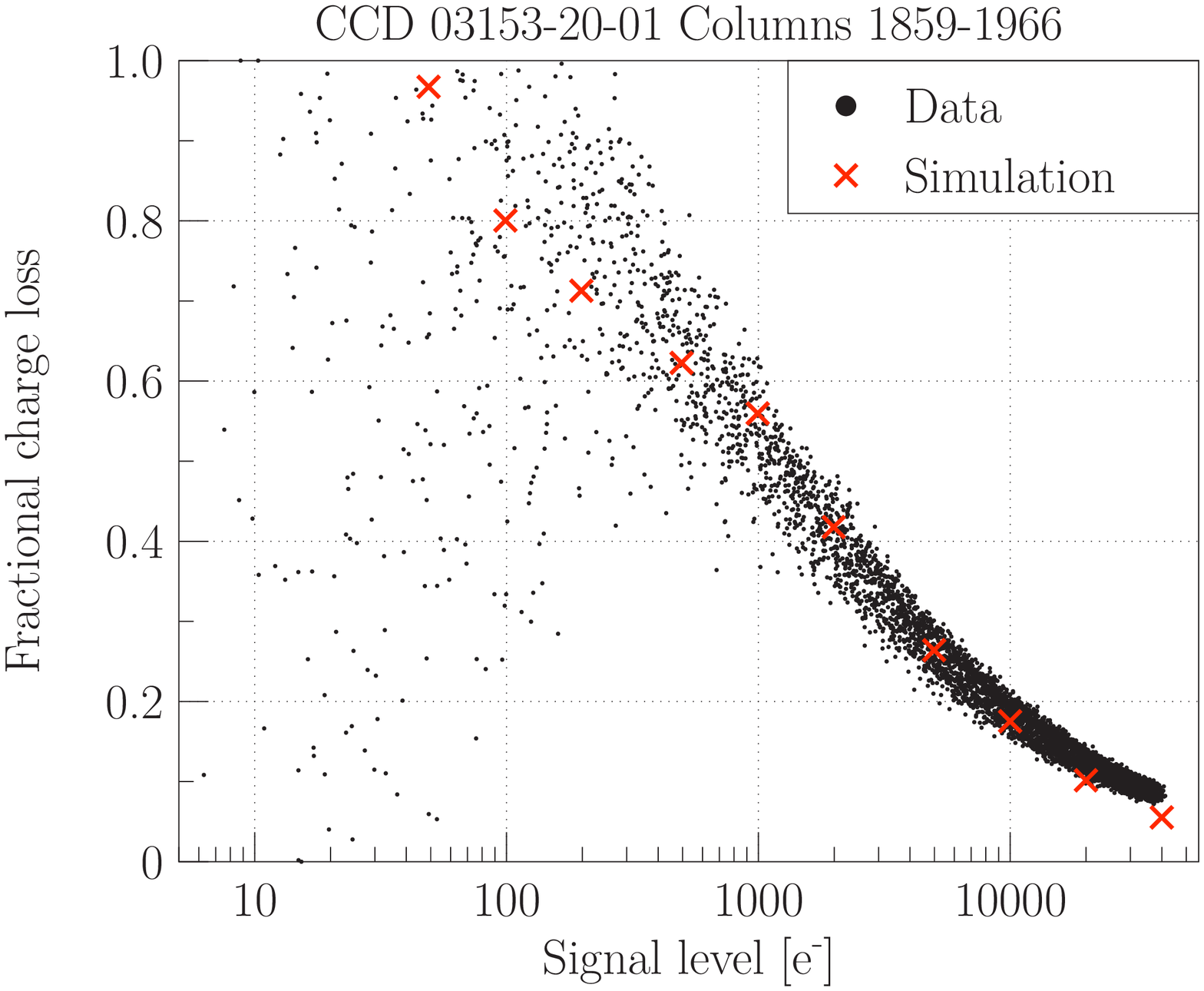}
  \includegraphics[width=0.49\textwidth]{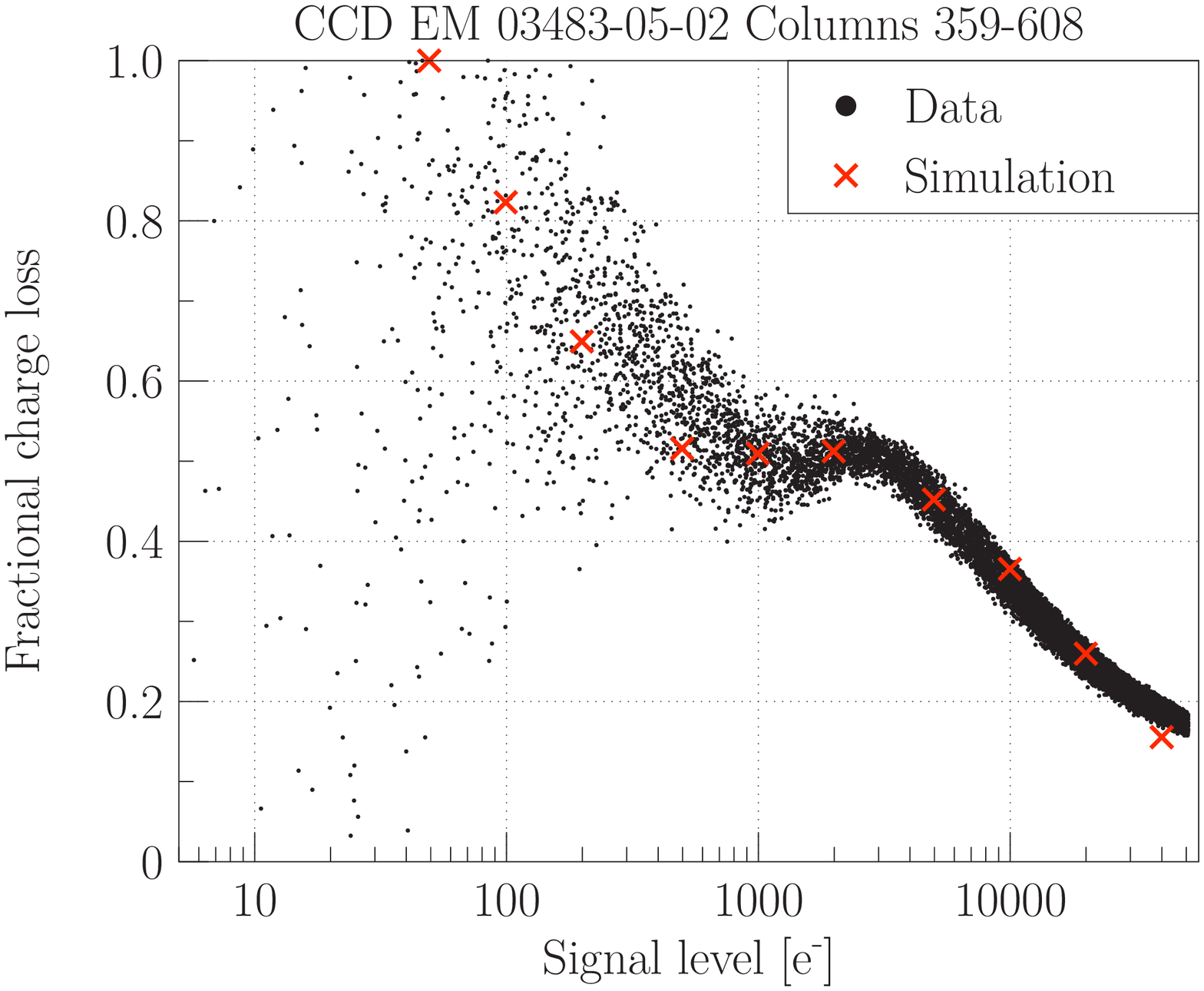}
  \caption{Comparison between experimental (black dots) and simulated (red crosses) FPR measurements as a function of CI level. The left panel shows Hopkinson's
  measurements of AF DM 03153-20-01 for which the upper half is suspected to have reduced SBC FWCs. The right
  panel shows a Hopkinson's measurements of AF EM 03483-05-02, which demonstrates typical SBC FWCs in both CCD halves. The difference in charge loss is due to different irradiation levels. The detailed Monte-Carlo model \citep{prodhomme2011}
  in combination with the analytical representation of the electron density distribution is able to
  qualitatively reproduce the FPR in both cases.  
  }
  \label{f:fitexamples}
\end{figure*}

\begin{figure*}
  \centering
  \includegraphics[width=0.49\textwidth]{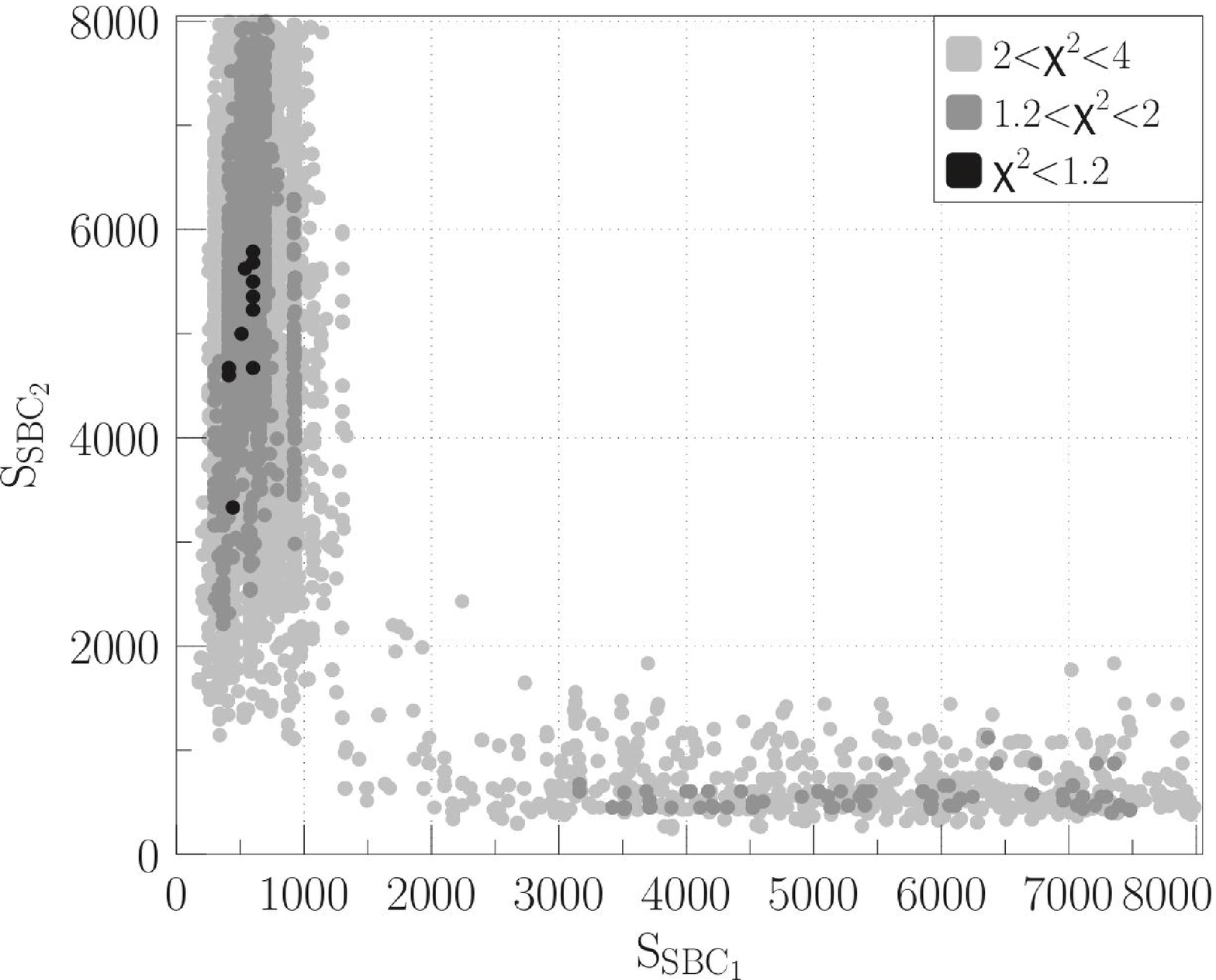}
  \includegraphics[width=0.49\textwidth]{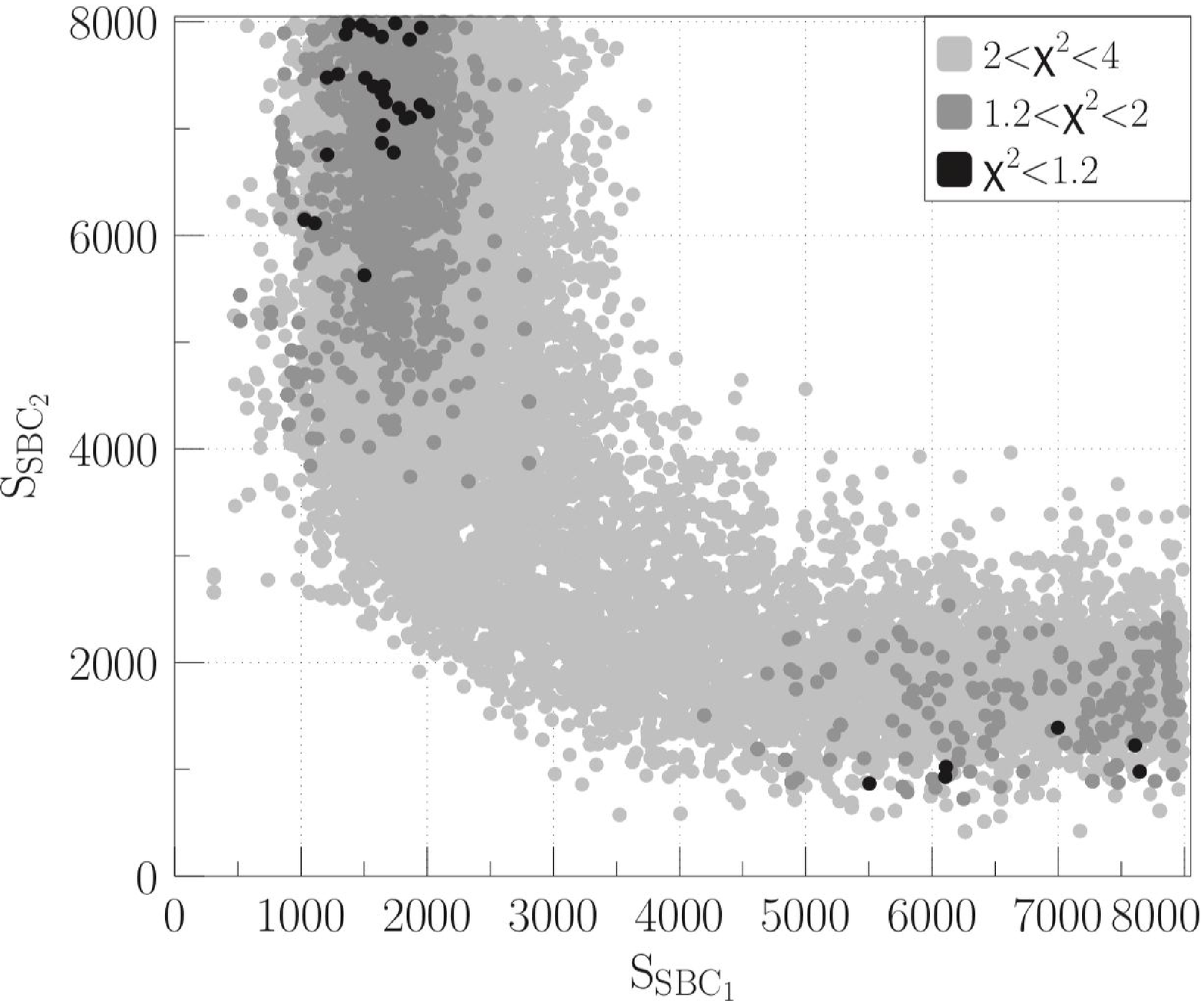}
  \caption{Maps of the resulting agreement between simulated and experimental FPR for devices suspected of containing reduced SBC FWCs (left) and typical SBC FWCs (right) as a function of $S_{\textrm{SBC}_1}$ and $S_{\textrm{SBC}_2}$, respectively the SBC FWC in the CCD upper half and lower half.
  The plotted agreement corresponds to the described $\chi^2$ (see Section~\ref{s:expt:sim:comparison}) normalized by the best $\chi^2$ achieved.
  The left panel shows the resulting map for the AF DM 03153-20-01 device (Fig.\ \ref{f:fitexamples} left), while the right panel shows the obtained map for the AF EM 03483-05-02 device (Fig.\ \ref{f:fitexamples} right).}
  \label{f:fitmap}
\end{figure*}

Fig.~\ref{f:fitexamples} shows the fractional charge loss measurement (first line of the CI considered only) carried out by
Hopkinson (black dots) on the {\gaia} AF CCD EM 03153-20-01 for the columns 1859 to 1966 (left panel) and on the {\gaia} AF CCD EM 03153-16-02 for the columns 609 to 858 (right panel) along with the result of the
presented simulation setup that aimed to reproduce these measurements (red crosses). The simulations shown
are representative of the best fit to the data that can be obtained following a semi-random search\footnote{A genetic algorithm was used as an optimization procedure - see\\\url{http://watchmaker.uncommons.org/}}  in the
parameter space for both investigated cases: suspected reduced SBC FWCs (left) and a typical SBC FWC case (right).

Fig.~\ref{f:fitmap} shows that the fit for the two cases occupies different regions of parameter space, as expected given the differences in Fig.~\ref{f:fitexamples}. The most likely solutions are where $S_{\mathrm{SBC}_1} < 3000$~e$^-$ or $S_{\mathrm{SBC}_2} < 3000$~e$^-$ for the typical SBC FWC case but where $S_{\mathrm{SBC}_1} < 1000$~e$^-$ or $S_{\mathrm{SBC}_2} < 1000$ e$^-$ for the reduced SBC FWC case.  From the figures it also appears that there are more solutions with low $\chi^2$ for $S_{\mathrm{SBC}_1} < S_{\mathrm{SBC}_2}$. In principle the two cases should be distinguishable because it matters which SBC FWC comes first. Of course with the data available the distinction is not so easy to make.  However, as $S_{\mathrm{SBC}_1}$ should always be smaller than $S_{\mathrm{SBC}_2}$, we consider the $S_{\mathrm{SBC}_2} < 3000$~e$^-$ solutions in both cases to be unphysical.  


 The first verification that SBCs are present in \gaia~CCDs by finding the characteristic bump in the FPR curve was also the first published estimate of {\gaia}'s SBC FWC \citep{hopkinson2005}.  Fig.~\ref{f:fitexamples} (right) plots includes the same data as their fig.~11.  They  
 estimated the SBC FWC to be $\sim$1400~\electron.  However, this is a typographical error in their paper.  The arrow defining their SBC FWC in their fig. 11 is pointing to $\sim$1400~ADU, which corresponds to $\sim$1100~\electron. Fig. ~\ref{f:fitexamples} (right) shows that this signal level
corresponds to the first inflexion point.  This paper is the first time a FPR curve has been modelled taking into account the different SBC FWCs in a \gaia~CCD.  $S_{\mathrm{SBC}_1} \approx 1500$~e$^-$ in Fig.~\ref{f:fitmap} (right) corresponds to a point in between the inflexion points in Fig.~\ref{f:fitexamples} (right), whereas $S_{\mathrm{SBC}_2} \approx 6000$~e$^-$ does not correspond to any part of the inflexion.  Therefore, the characteristic bump is at signal levels that correspond to the smaller upper half SBC FWC and not the larger lower half SBC FWC.
 
 Because of the way e2v manufacture the CCDs (all the SBC doping
implanted simultaneously), intra-stitch block SBC FWCs in a particular half should be the
same. However, in the {\gaia} case, there is no guarantee that inter-stitch block SBC FWCs will be
the same. This is due to the SBC doping abutting the ABD doping (see
Fig.~\ref{f:pixel} top schematic) and overlapping the ABD shielding doping (see Fig.~\ref{f:pixel}
middle schematic). The ABD doping is the first pixel feature to be implanted into each pixel within
a stitch block area of CCD silicon using its own photo-lithographic mask (the size of a stitch
block). The ABD mask is aligned to the `zero grid'. All the SBCs in all stitch blocks have their
doping implanted subsequently using a SBC photo-lithographic mask aligned to the same zero grid. Each
mask alignment to the zero grid is subject to random alignment (stitch) errors of $\le$0.25 $\mu$m
\citep{burt2005}. Because the ABD shielding doping cancels out the SBC doping, the effective SBC
doping width, consequent potential depth and resulting FWC is sensitive to the stitch error.
Therefore, scatter in inter-stitch block SBC FWC (both intra-CCD and inter-CCD) is expected from the
nominal e2v design of the {\gaia} CCDs, as seen in Fig. \ref{f:3egs}.   

The SBC FWCs of the typical case are much smaller than e2v's analytical {\gaia} CCD design predictions (published for the first time in
\citealt{seabroke2010}): 7900 \electron in the upper half and $13\,000$ \electron in the lower half, which correspond to $(w_{\textrm{eff,1}},w_{\textrm{eff,2}}) = (3,4)$ $\mu$m.  Inputting $(w_{\textrm{eff,1}},w_{\textrm{eff,2}}) = (3,4)$~$\mu$m to the \cite{seabroke2010} pixel model successfully reproduces e2v's analytical predictions (see Fig. \ref{f:weff}).  This figure suggests this is not due to random stitch errors (worst-case
random alignment stitch error = $-$0.5 $\mu$m) as these would result in $(w_{\textrm{eff,1}},w_{\textrm{eff,2}}) = (2.5,3.5)$~$\mu$m, which does not reduce the SBC FWCs to the model fit of  $(S_{\mathrm{SBC}_1},S_{\mathrm{SBC}_2}) = (1509,7474)$~\electron.  To do this requires a systematic shift to $(w_{\textrm{eff,1}},w_{\textrm{eff,2}}) \approx (2.00,2.75)$~$\mu$m\footnote{$w_{\textrm{eff,2}} \approx 2.75$~$\mu$m is larger than $y_{\mathrm{SBC},\mathrm{max}} = 2$~$\mu$m.  Nevertheless, Fig.~\ref{f:fitexamples} (left) shows that the simulation is able to reproduce the FPR data very well.}.  This $\Delta(w_{\textrm{eff,1}},w_{\textrm{eff,2}}) \approx (1.00,1.25)$~$\mu$m from the design appears to be uncalibrated systematic offsets in e2v photo-lithography, which could either be
due to systematic stitch offsets or ABD shield doping diffusion in the AC direction. The \cite{seabroke2010}
pixel model does not specifically simulate either of these scenarios, rather it simulates a $w_{\textrm{eff}}$ that could be produced by both or either of these scenarios.


The best fit in Fig.~\ref{f:fitmap} (left) is $(S_{\mathrm{SBC}_1},S_{\mathrm{SBC}_2}) = (442,3331)$~\electron, which are indeed reduced SBC FWCs compared to the typical case by a factor of 3 in the upper half and a factor of 2 in the lower half.  This corresponds to $(w_{\textrm{eff,1}},w_{\textrm{eff,2}}) = (1.75,2.25)$~$\mu$m.  The difference with the typical case is $\Delta(w_{\textrm{eff,1}},w_{\textrm{eff,2}}) \approx (0.25,0.50)$~$\mu$m.  Because this AL stitch block couple consists of termination stitch blocks, it is possible that this is a rare example of compounding stitch errors starting from the typical SBC FWC case.

\section{Pocket pumping measurements} \label{s:pocket} 

\subsection{Technique} 

The determination of SBC FWCs in both halves of \gaia~CCDs in the previous section required simulations to interpret measurements of charge packets that had traversed both halves of the CCD (FPR) in terms of SBC FWC in each half. Pocket pumping can provide a direct
measurement of the SBC FWC in each half. Charges from a flat field illumination are moved back
and forth in the image area over a one pixel length. In pixels with traps, single electrons can be
statistically trapped during one half-cycle and then released during the next half-cycle into the
adjacent pixel. This produces bright-dark pairs around the mean flat field level at the trap
position. Repeating the technique with increasing mean flat field signal levels identifies the
number of traps per pixel as a function of signal size. 

\subsection{Devices tested} 

\begin{figure}
  \centering
  
  \includegraphics[width=0.3\textwidth]{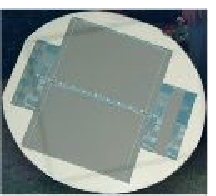}
  \caption{e2v {\gaia} silicon wafer with photoactive regions (dark grey) showing two CCD91-72s in the centre and small test structures on either side.  The photoactive strip on the right side is two CCD221s abutted.  (Image courtesy of e2v technologies.)}
  \label{f:wafer}
  
  \includegraphics[width=0.35\textwidth]{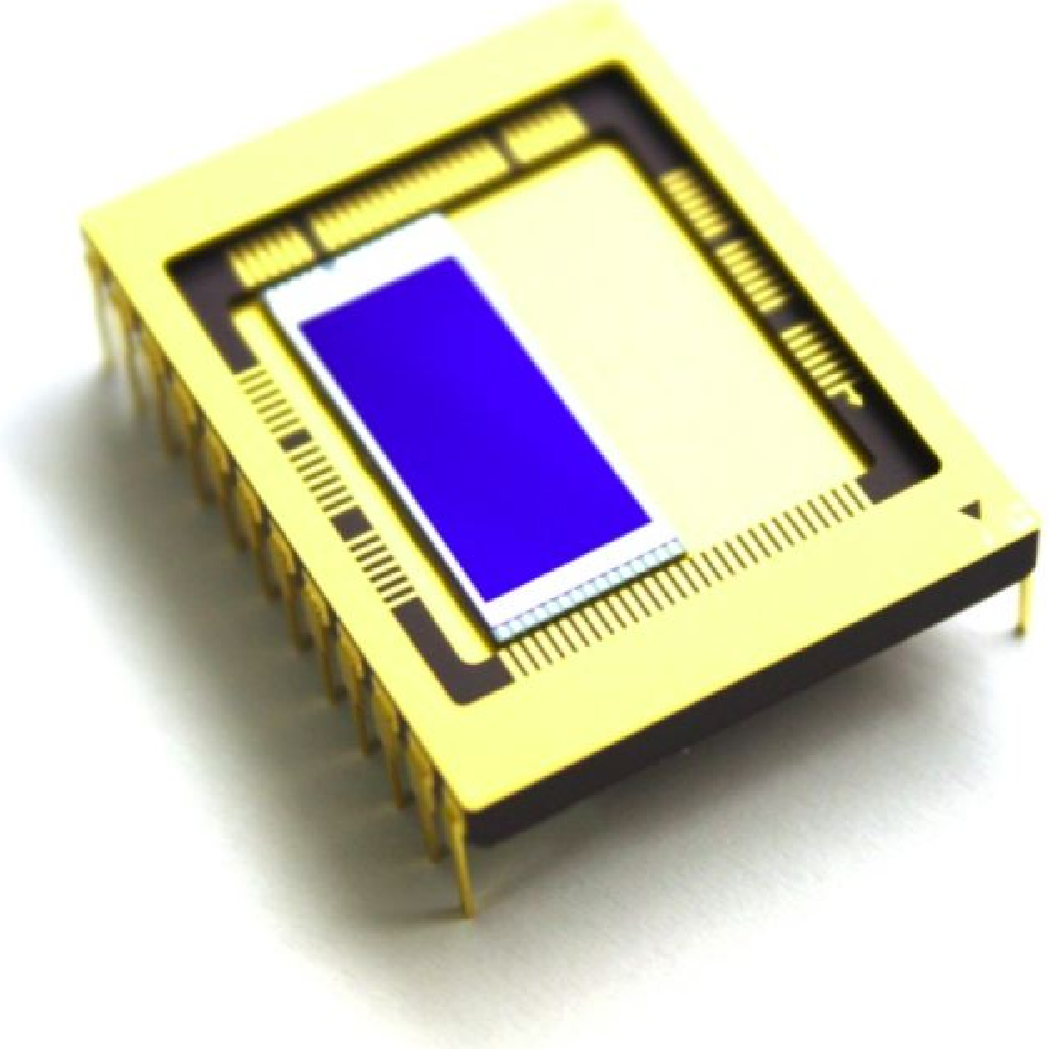}
  \caption{{\gaia} test structure CCD221 packaged for use. (Image courtesy of e2v technologies.)}
  \label{f:ccd221}
  
  \includegraphics[width=0.47\textwidth]{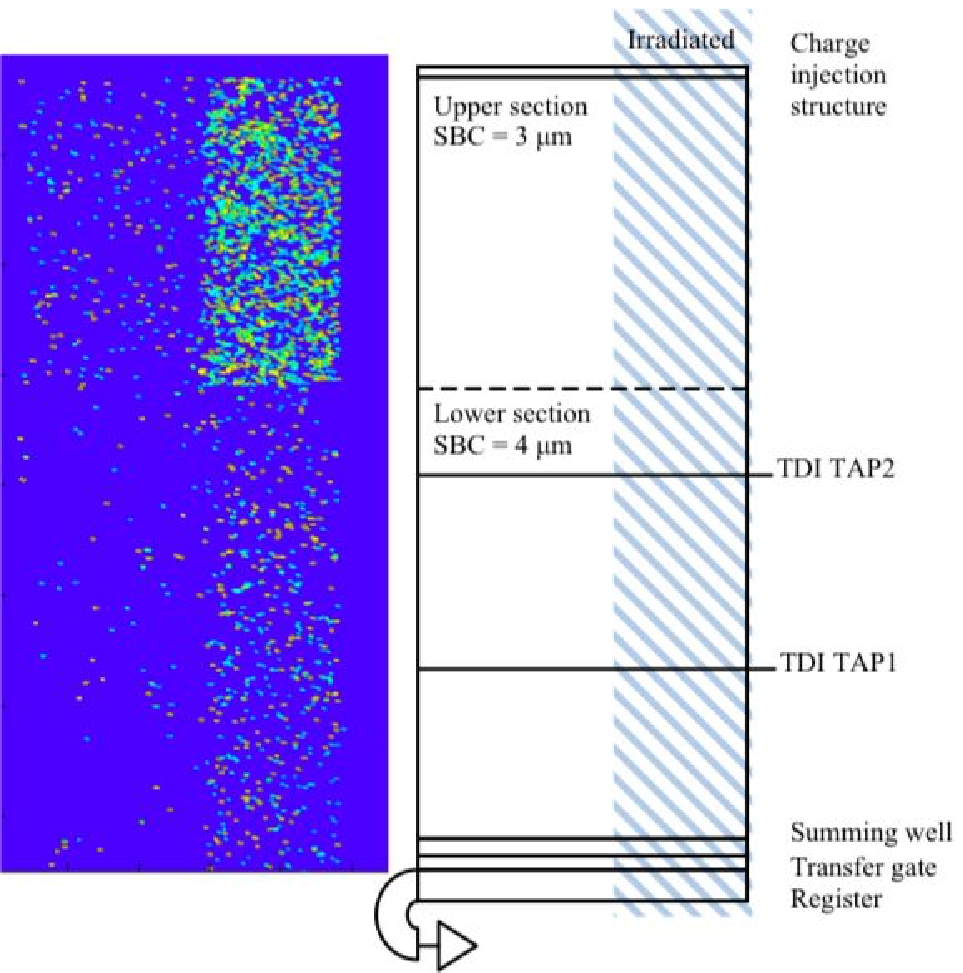}
  \caption{{\it Left:} Image of a pocket pumping measurement of test structure CCD221 (serial number 06026-16-04) showing forward (yellow dots) and reverse (cyan dots) traps.  {\it Right:} Layout of all CCD221 test structures and the area of 06026-16-04 that was subjected to proton irradiation to a 10 MeV equivalent dose of $1\times 10^9$ protons cm$^{-2}$.}
  \label{f:ccd221layout}
\end{figure}

Fig. \ref{f:wafer} shows that each e2v {\gaia} silicon wafer can only include two {\gaia} CCDs (CCD91-72).  The room left on each wafer was used for test structures that were included to assess the radiation hardness of {\gaia} CCDs.  In order for these test structures to be irradiated and tested, they have to be packaged.   e2v packaged up two CCD221 test structures for the e2v centre for electronic imaging at The Open University to test (see Fig. \ref{f:ccd221}).  Because they have the same pixel architecture (and thus manufacturing processes) as {\gaia} CCDs, we have repeated the pocket pumping measurements conducted by \citet{kohley2009} (hereafter K09) to test whether these test structures also have reduced SBC FWCs.   K09 tested an AF FM CCD with serial number 05486-11-02.  The test structures are also post-2004 but from two different batches: 06026-16-04 and 06095-09-04.  The last two digits are the position of the device on the wafer: 03 and 04 are the two right side positions for the CCD221 test structures (see Fig. \ref{f:wafer}).  06026-16-04 was irradiated (see Fig. \ref{f:ccd221layout}).  06095-09-04 was not irradiated.

All CCD221 test structures have 1440 TDI lines and 224 columns, which is smaller than one {\gaia} stitch block (2160 or 2340 TDI lines and 250 or 108 columns).  The CCD221 is not stitched during manufacture i.e. its image area consists of only one stitch block but the SBC doping has a width change (at the dotted line in Fig. \ref{f:ccd221layout}) to replicate the width transition at the AC stitch boundaries in CCD91-72's.  This means the CCD221 test structures cannot be used to compare stitch blocks in different AC positions as was done in K09. 

\subsection{Results}\label{s:pp_results} 

\begin{figure}
  \centering
  \includegraphics[width=0.49\textwidth]{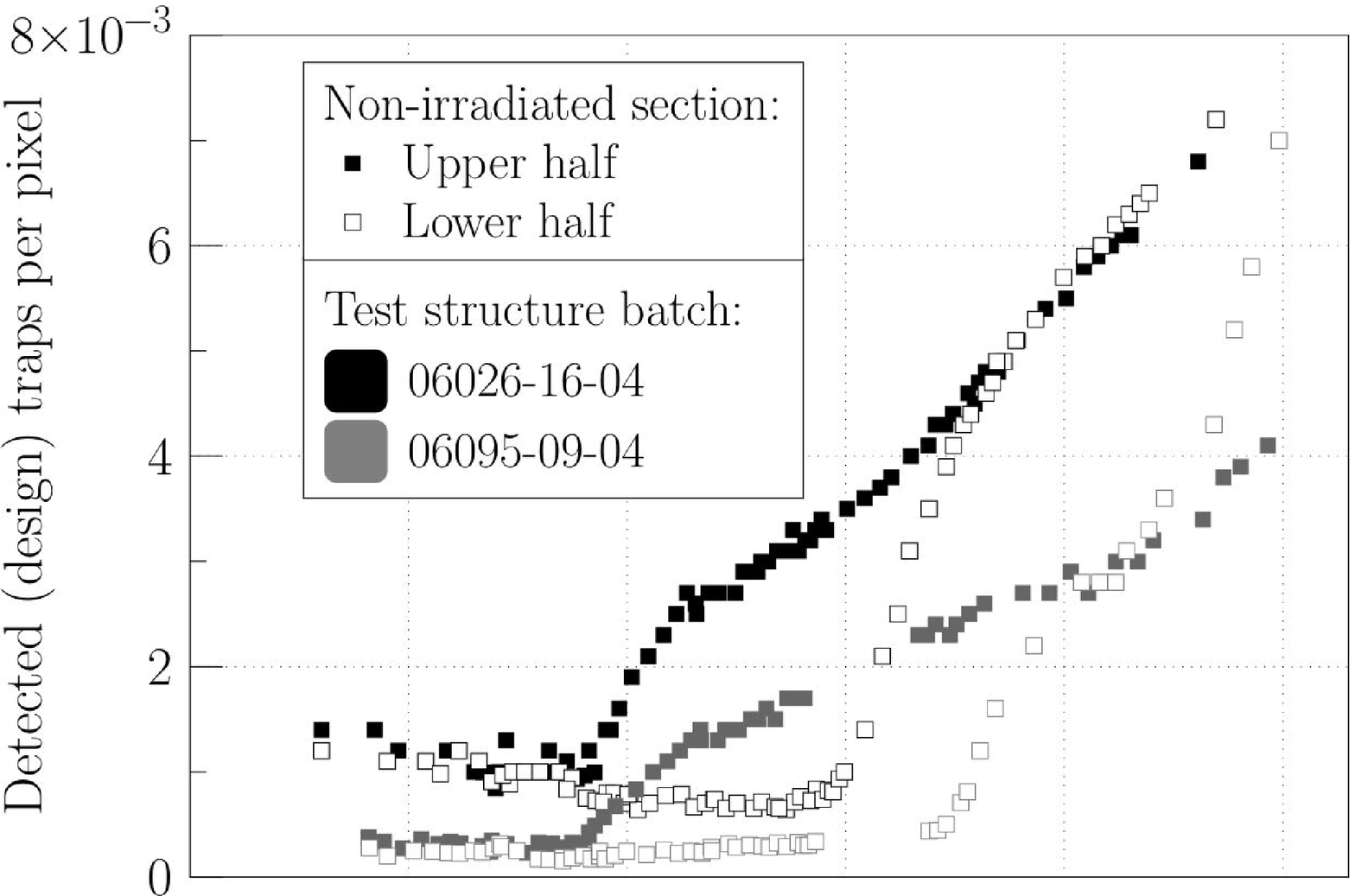}
  \includegraphics[width=0.49\textwidth]{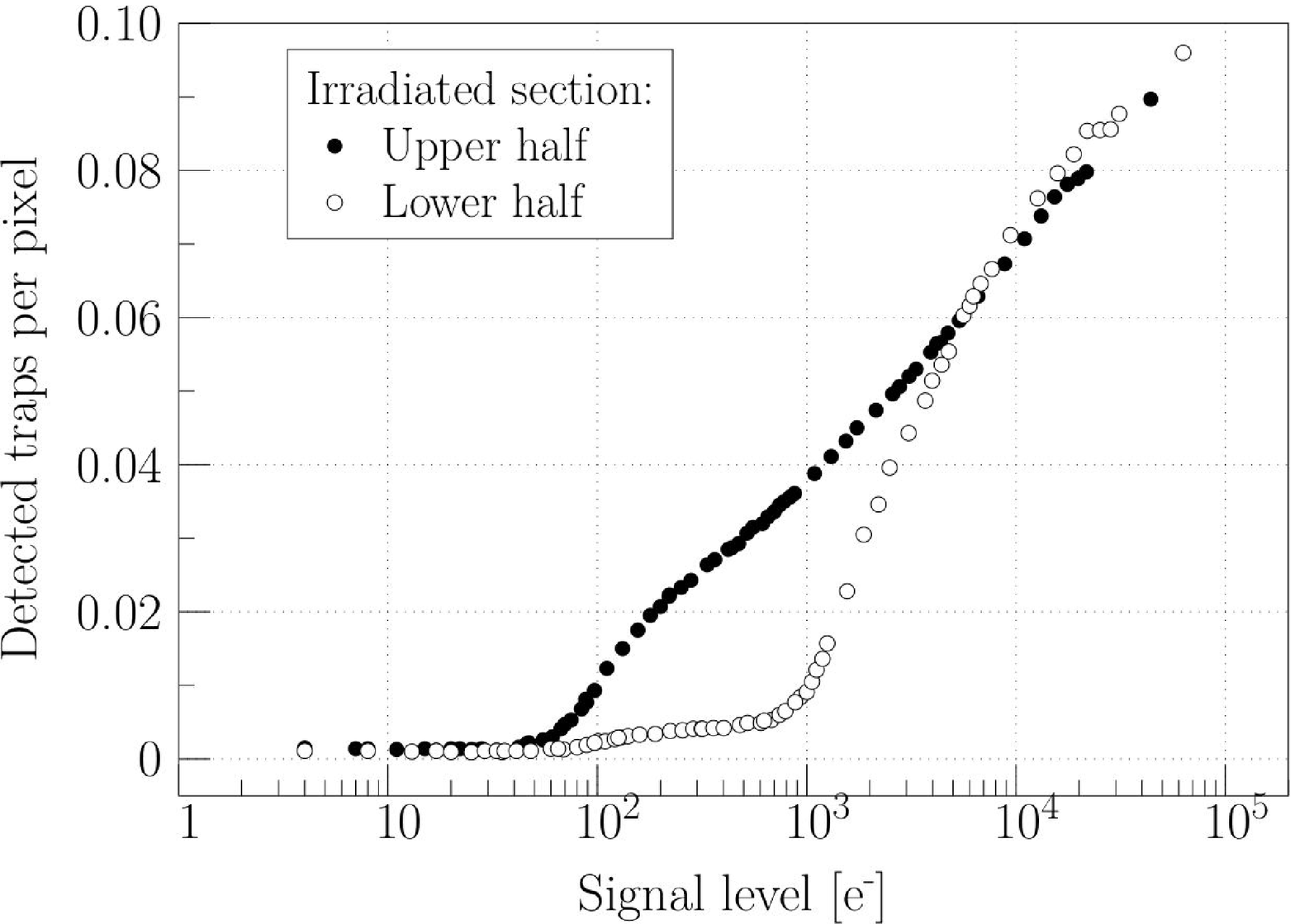}
  \caption{Number of detected traps per pixel as a function of signal level. {\it Top:} Fabrication traps in both test structures (see left side of pocket pumping image and left side of layout schematic in Fig.~\ref{f:ccd221layout}).  {\it Bottom:} Radiation traps in 06026-16-04 (see right side of pocket pumping image and right side of layout schematic in Fig.~\ref{f:ccd221layout}).}
  \label{f:fabrad}

  \includegraphics[width=\columnwidth]{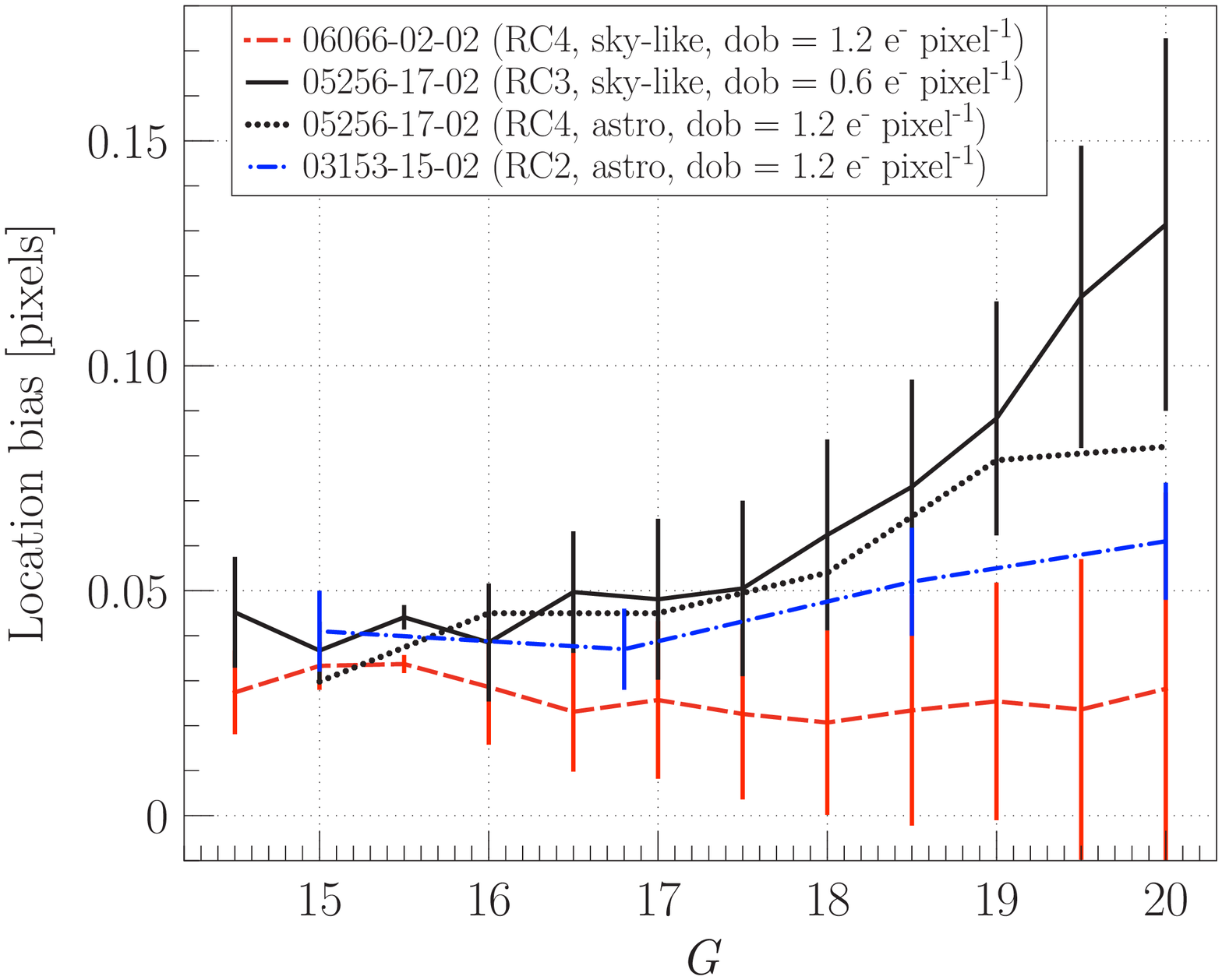}
  \caption{Stellar image location bias values calculated through the analysis of data from a
  partially irradiated CCD using a mask with pseudo sky-patterns of pin-holes and CI interval of 1 s.}
  \label{f:rc_bias}
\end{figure}

The top of Fig. \ref{f:fabrad} shows that the number of identified fabrication traps per pixel in both test structures is the same order of magnitude as seen in by K09 (see their fig. 9).  It also shows that there are an order of magnitude more radiation traps per pixel as there are fabrication traps per pixel.  Nevertheless, the same trends as a function of CCD half are seen in both the non-irradiated and irradiated parts of 06026-16-04.  The lower half trend as a function of signal size in this device is that signals $<$$1000$~\electron~have a very different gradient to signals $>$$1000$~\electron.  The lower half trend as a function of signal size in the other device (06095-09-04) is similar: signals $<$$3000$~\electron~have a very different gradient to signals $>$$3000$~\electron.  We interpret the change in gradient as due to electrons spilling out of SBCs in the lower halves of these devices with FWCs of $\approx$1000 and 3000~\electron respectively.  K09 found exactly the same spread of SBC FWCs between the AL stitch block couples in their CCD.  It suggests that both test devices and the K09 CCD all have $w_{\textrm{eff,2}}$ between 1.75 and 2.25~$\mu$m (see Fig. \ref{f:weff}).

Like K09's fig. 9, the upper halves of our test structures do not exhibit a change in gradient at $\approx$$1000$~\electron.  Instead the upper half trend as a function of signal size in both devices is that signals $<$$50$~\electron~have a very different gradient to signals $>$$50$~\electron.  We interpret the change in gradient as due to electrons spilling out of SBCs in the upper halves of these devices with FWCs of $\approx$50~\electron.  K09's smallest signal level of $40$~\electron did not find the change in gradient, suggesting the upper half SBC FWC of this CCD is $<$40~\electron.  This is consistent with our findings because even a sample random stitch error can result in a large change in SBC FWC because the SBC FWC-$w_{\textrm{eff}}$ relation is steep at this signal level.  Such small SBC FWCs may not be physical because the SBC potentials would be very shallow, allowing electrons to thermally diffuse out.  Whether the upper half SBC FWCs in these devices is zero or tens of electrons, these SBC FWCs are very much reduced compared to their lower halves and compared to pre-2004 CCDs. 

The post-2004 devices have $(w_{\textrm{eff,1}},w_{\textrm{eff,2}}) = (<1.5,2.0)$~$\mu$m, which is different to the typical pre-2004 devices with $(w_{\textrm{eff,1}},w_{\textrm{eff,2}}) = (2.00,2.75)$~$\mu$m ($\Delta(w_{\textrm{eff,1}},w_{\textrm{eff,2}}) \approx (>0.5,0.75)$~$\mu$m) and very different to the design with $(w_{\textrm{eff,1}},w_{\textrm{eff,2}}) = (3,4)$~$\mu$m ($\Delta(w_{\textrm{eff,1}},w_{\textrm{eff,2}}) \approx (>1.5,2)$~$\mu$m, see Fig. \ref{f:weff}).  Again, this appears to be uncalibrated systematic offsets in e2v photo-lithography but larger in magnitude than the pre-2004 offsets.

%
%

\section{Comparison of EADS-Astrium Radiation Campaigns 3 and 4 test data from the same CCDs}\label{s:discuss:rc}


EADS-Astrium have carried out a series of five test
campaigns on a small number of irradiated CCDs in order to evaluate the impact of radiation damage
on the image parameter estimation and ultimately on the {\gaia} astrometric performances.
Table~\ref{t:rc_CCDs} provides a summary of the CCDs tested during these campaigns. 
The tests were carried out using a number of runs of an illuminated mask along the CCD which is operated in TDI mode. A number of
different masks were used throughout the testing campaigns, each with a different pattern of pin-holes (of sometimes differing diameters). The location and flux biases  are generally calculated through a comparison of data acquired from an irradiated portion of the  CCD with an non-irradiated portion of the same device.


The resulting bias measurements shown in Fig.~\ref{f:rc_bias} differ between RC2, RC3 and RC4 at
faint magnitudes: the RC3 data does not plateau to an approximately constant bias at faint magnitudes.
This is in contrast to the data from the other campaigns. One possible reason for the RC3 bias
values not plateauing is due to the presence of the constant Diffuse Optical Background (DOB) being
better controlled  in the experimental setup of RC3 compared to RC2 and RC4. When the DOB is very
close to zero, then just a small amount of background light can drastically reduce the image location bias
due the resulting photo-electrons keeping many of the traps with long time constants (or `slow' traps) filled.
Indeed, the measured DOB levels are 0.6 $\electron$ pixel$^{-1}$ lower in the RC3
data than in RC4. However, RC4 data taken with the same CCD as was used in RC3 shows a similar
bias function to RC3 (i.e., no plateau at faint magnitudes) and in this case the DOB level is comparable
to both the RC2 and RC4 data set with another CCD. It thus appears that the DOB level is not the primary cause of the
discrepancies between the location bias functions across the different test campaigns.   This hypothesis is further strengthened when one considers that these data were obtained with a CI implemented every second, which also has the effect of keeping the  `slow' traps filled, and making the effect of small levels of  DOB much less significant. 

The most likely conclusion is that the measured differences are intrinsic to the CCDs.  Fig.~\ref{f:rc_bias} shows that 03153-15-02's location bias as a function of brightness is in between the plateau of 06066-02-02 and 05256-17-02. \cite{prodhomme2011} modelled the FPR curve in different AL stitch block couples in 03153-15-02 from a single non-uniform CI using a single SBC FWC for both halves of the CCD.  They found the SBC FWC to be 2825 electrons.  Given that Section  \ref{s:expt:results} suggests that the shape of the FPR curve is dominated by the upper half SBC FWC, we can assume the SBC FWC in the upper half of 03153-15-02 is approximately 2825 electrons.  The typical pre-2004 case in Section  \ref{s:expt:results} has SBC FWCs of 1509 electrons.  Both this value and 2825 electrons are consistent with $w_{\textrm{eff,1}} \approx 2.00 \pm 0.25$~$\mu$m, which corresponds to $S_{\textrm{SBC,1}} \approx 2000 \pm 1000$~\electron (see Fig. \ref{f:weff}).  Therefore, we also consider 03153-15-02 to be a typical pre-2004 case and so not have reduced SBC FWCs.

The SBC FWCs of 06066-02-02 and 05256-17-02 have not been measured.  Nevertheless, the fact that Fig.~\ref{f:rc_bias} shows that the location bias of 06066-02-02 is smaller than 03153-15-02 suggests that the 06066-02-02 SBC FWCs are larger than 03153-15-02's because they are protecting small charge packets from CTI more than 03153-15-02.  This is the first evidence that not all post-2004 CCDs have reduced SBC FWCs compared to the typical pre-2004 CCD.   Following the same argument, the fact that Fig.~\ref{f:rc_bias} shows that the location bias of 05256-17-02 is larger than 03153-15-02 suggests that the 05256-17-02 SBC FWCs are smaller than 03153-15-02's because they are protecting small charge packets from CTI less than 03153-15-02.  However, whether 05256-17-02 SBC FWCs are the same as all the other post-2004 SBC FWCs measured in Section \ref{s:pocket} is an open question.  

\begin{table}
  \centering 
  \caption{Summary of EADS-Astrium's RC CCDs.}
  \begin{tabular}{lllll}
    \hline
    Serial Number & Variant  & Reduced SBC FWCs				\\
    \hline
    03153-15-02 & AF DM & No 		\\
    05256-17-02	 & AF EM & Yes?  \\
    06066-02-02	 & AF EM & No? 	\\
    06244-03-01	 & RP DM & ? 	\\
    06273-08-01	 & RP EM &  ? \\
    \hline
  \end{tabular}
  \label{t:rc_CCDs}
\end{table}

\section{Discussion} \label{s:discuss}


%
%
\subsection{Comparison of pre- and post-2004 CCD samples}\label{s:discuss:stats}



We consider evidence from the pre-2004 and
post-2004 CCDs separately because FM CCDs only will be flown on {\gaia} and all FM CCDs were built post-2004.  Because of the low number of CCDs that have been tested, we augment our evidence found in this paper with evidence from the literature. 

\subsubsection{Pre-2004 CCDs}

 From the Sira test sample, out of the seven 2003 AF CCDs searched for
evidence of reduced SBC FWCs in Section \ref{s:expt}, we only find one CCD with one stitch block AL couple (termination stitch block with 108 columns) strongly suspected of
not having reduced SBC FWCs compared to the typical case (by one order of magnitude in the upper stitch block).   We augment  this pre-2004 CCD sample with the first device tested by
EADS-Astrium, an AF DM (03153-15-02) that exhibits typical SBC FWCs in upper half stitch block couples. Thus, a reduction in upper half SBC FWCs by one order of magnitude affects the pre-2004 CCD sample as follows:
\begin{itemize}
\item 1 out 8 CCDs $= 12.5$\% of CCDs;
\item 1 out of ($9\times8=$) 72 stitch block AL couples in 8 CCDs $\approx$1.4\% of stitch block AL couples;
\item 108 out of [8 $\times(2\times108 + 7 \times 250) =$] $15\,728$ columns in 8 CCDs $\approx$0.7\% of columns.
\end{itemize}  

This small number of columns with reduced SBC FWCs in their upper halves is consistent with accumulating stitch errors at each of the eight inter-stitch block AC alignments being a rare event, mainly in the termination stitch blocks furthest from the readout node.

\subsubsection{Post-2004 CCDs} 

In Section \ref{s:pocket}, we found two CCD221 test structures (224 columns in each device) with reduced SBC FWCs in their upper stitch blocks compared to the typical pre-2004 case (by two orders of magnitude).   To these we add the device tested by \cite{kohley2009} (where all 9 stitch block AL couples were tested). Thus, a reduction in upper half SBC FWCs by two orders of magnitude affects the post-2004 CCD sample (that have had their SBC FWCs measured) as follows: 
\begin{itemize}
  \item 3 out 3 CCDs $= 100$\% of CCDs;
  \item ($1+1+9=$) 11 out of ($1+1+9$) 11 stitch block AL couples tested in 3 CCDs
    $\approx$100\% of stitch block AL couples;
  \item ($224+224+1966=$) 2454 out of ($224+224+1966=$) 2454 columns tested in 3 CCDs
    $\approx$100\% of columns.
\end{itemize}

This suggests that a rare compounding of stitch errors is not responsible for the reduction of SBC FWCs in the post-2004 sample but a systematic effect in e2v photo-lithography.  Posing the more general question of how many post-2004 devices show any direct or indirect evidence for a reduction in SBC FWCs compared to the typical pre-2004 SBC FWCs allows us to include the two post-2004 devices have been tested by EADS-Astrium in their RCs:

\begin{itemize}
  \item 4 out 5 CCDs $= 80$\% of CCDs;
  \item ($1+1+2+9=$) 13 out of ($1+1+2+5+9$) 18 stitch block AL couples tested in 5 CCDs
    $\approx$72\% of stitch block AL couples;
  \item ($224+224+500+1966=$) 2914 out of ($224+224+500+1250+1966=$) 4164 columns tested in 5 CCDs
    $\approx$70\% of columns.
\end{itemize}

\subsubsection{Comparison conclusion}

It is clear that the pre- and post-2004 CCD samples are systematically different, despite the small sample sizes.  The systematic difference is in the typical SBC FWCs for each sample.  The pre-2004 sample typically has upper half SBC FWCs of thousands of electrons, with only one exception of 400 electrons.  The post-2004 sample typically has upper half SBC FWCs of tens of electrons (or zero), with only one exception of indirect evidence of thousands of electrons.  
This points to a change in e2v manufacturing of {\gaia} CCDs between 2003 and 2005.  
A different mask set is the only known change in e2v manufacturing of {\gaia} CCDs between 2003 and 2005.  However, this can only be considered as circumstantial evidence to explain any systematic changes between pre- and post-2004 CCDs because it would be very difficult to definitively prove the mask set was responsible due to a plethora of other factors that go into manufacturing CCDs with complex pixel architectures.  

  In the introduction we asked the question: are the SBC FWCs in {\gaia} CCDs used to predict the faint star ($13 \le G \le 20$ mag) astrometric performances
    and against which mitigation models are being developed and tested representative of the CCDs that will fly on \gaia? 03153-15-02 was used for these activities.  However, this paper has shown its upper half SBC FWC to be typical of the pre-2004 sample.  The pre-2004 sample does not appear to be representative of the CCDs that will fly on \gaia~(the post-2004 sample).  This means the SBC FWCs of 03153-15-02 also may not be representative of the actual SBC FWCs on \gaia.  The 03153-15-02 SBC FWCs were used in the image location estimation in the presence of radiation damage in \cite{prodhomme2011b}, which were used in \cite{holl2012} to investigate the effect of image location errors on the astrometric solution.  Hence, it is not known how representative of the whole focal plane these predictions are. 

\subsubsection{Predicting the SBC FWCs of \gaia's onboard CCDs}

SBC fabrication occurs during front-face processing, which is when the electrodes are made and all doping implanted and is batch based.  Because all the {\gaia} CCDs have the same pixel architecture, front-face processing is the same for all {\gaia} CCDs: AF, Blue Photometer (BP), Red Photometer (RP). e2v and EADS-Astrium refer only to these CCD variants as the Radial Velocity Spectrometer (RVS) CCDs are identical to RP CCDs. Each CCD serial number encodes when the front-face processing occurred.  The fact that all wafers in a batch have the same nominal front-face processing means that if a CCD of any type is found to be affected by reduced SBC FWCs, then it is possible that all the CCD variants (AF, BP and RP) from that batch may also be have reduced SBC FWCs.  Table \ref{t:fm} summarizes all the post-2004 CCDs from which SBC FWCs have been derived and their batch numbers.  It shows they are all from different batches so whether other CCDs from the same batch are also affected remains unproven (none of the other CCDs that could be tested are from these four batches either).  In the absence of evidence demonstrating that all CCDs from the same batch have the same SBC properties, we cannot attempt to predict the SBC FWCs of CCDs onboard the \gaia~satellite (post-2004 FMs). 

\begin{table*}
  \centering 
  \caption{Summary of all the post-2004 CCDs from which SBC FWCs have been derived: method (PP is pocket pumping and RC is analysis of Radiation Campaign data); serial and batch numbers; and SBC FWCs in the upper halves ($S_{\mathrm{SBC}_1}$) and lower halves ($S_{\mathrm{SBC}_2}$) in electrons. $x$? indicates the SBC FWC is not derived from direct measurements but inferred from indirect measurements.  ? on its own indicates unknown.}
  \begin{tabular}{llllll}
    \hline
    Method & Variant & Serial Number  & Batch Number & $S_{\mathrm{SBC}_1}$ & $S_{\mathrm{SBC}_2}$\\
    \hline
     RC3\&4 (this paper: re-analysis) & AF EM CCD & 05256-17-02 & 05256     & $<2825$? & ?\\
    PP (K09: 1st measurement \& analysis)  & AF FM CCD & 05486-11-02	& 05486	 & $<40$ & 1000-3000\\
   PP (this paper: 1st measurement \& analysis) & AF FM test structure & 06026-16-04  & 06026        & 50 & 1000\\
   RC4 (this paper: re-analysis) & AF EM CCD & 06066-02-02 & 06066 & $>2825$? & ?\\
   PP (this paper: 1st measurement \& analysis) & AF FM test structure & 06095-09-04 & 06095 & 50 & 3000 \\
        \hline
  \end{tabular}
  \label{t:fm}
\end{table*}

\subsection{Impact on the {\gaia} image location accuracy}\label{s:discuss:accuracy}

Given that we cannot predict the SBC FWCs of CCDs onboard the \gaia~satellite (post-2004 FMs), we simulate all the possible different SBC FWC scenarios to compare their potential impact on the \gaia~image location accuracy.  \cite{prodhomme2011b} fig. 23 shows that they have already simulated the no SBC scenario, ($S_{\mathrm{SBC}_1},S_{\mathrm{SBC}_2}) = (0,0)$, and the ``Functional SBC" scenario.  The latter uses a single SBC FWC derived from 03153-15-02:  $(S_{\mathrm{SBC}_1},S_{\mathrm{SBC}_2}) = (2825,2825)$ electrons.  In this paper we have found two other scenarios:

\begin{enumerate}
  \item The apparently rare pre-2004 scenario of one order of magnitude reduction of SBC FWCs compared to the typical pre-2004 case: $(S_{\mathrm{SBC}_1},S_{\mathrm{SBC}_2}) = (442,3331)$ electrons, $(\eta_1,\eta_2) = (0.132,0.073)$
  \item The apparently common post-2004 scenario of two orders of magnitude reduction of the upper half SBC FWCs compared to the typical pre-2004 case: $(S_{\mathrm{SBC}_1},S_{\mathrm{SBC}_2}) = (50,2825)$ electrons, $(\eta_1,\eta_2) = (0.132,0.1)$
\end{enumerate}

The simulations of image location accuracy use the same model \citep{prodhomme2011} as the simulations of the FPR curve in Section \ref{s:expt}, which includes the same parameterisation of the electron density distribution.  For scenario (i), the SBC FWC and electron density distribution spread factor ($\eta$) of both halves of the CCD are derived directly from the simulation of the FPR curve and and so form the input to simulate the image location accuracy.  In contrast, for scenario (ii) the SBC FWCs are not derived using the model so only the SBC FWCs are available as input to the model but not $\eta$.  Because the $S_{\mathrm{SBC}_1}$ in (i) is the closest in value to $S_{\mathrm{SBC}_1}$ in (ii), the $\eta_1$ in (i) is our best estimate for (ii) and so we use this.  We use the SBC FWC and $\eta$ values from \cite{prodhomme2011} as $S_{\mathrm{SBC}_2}$ and $\eta_2$ in scenario (ii) because they are similar to the lower half SBC FWCs found in Section \ref{s:pocket}.   


{\gaia} has been designed to perform absolute astrometric measurements at very high accuracy. As the
estimated image location for all CCD observations are ultimately used to derive the stellar astrometric
parameters, the requirements on the image quality are very stringent. Radiation damage distorts the image and
decreases the signal-to-noise ratio. While the image distortion, if not properly taken into account,
introduces a significant bias (e.g. Fig.~\ref{f:rc_bias}) in the image location estimation, the
decrease in signal-to-noise ratio implies an irreversible loss of accuracy independent of any
estimator. 

\begin{figure}
\centering
  \includegraphics[width=\columnwidth]{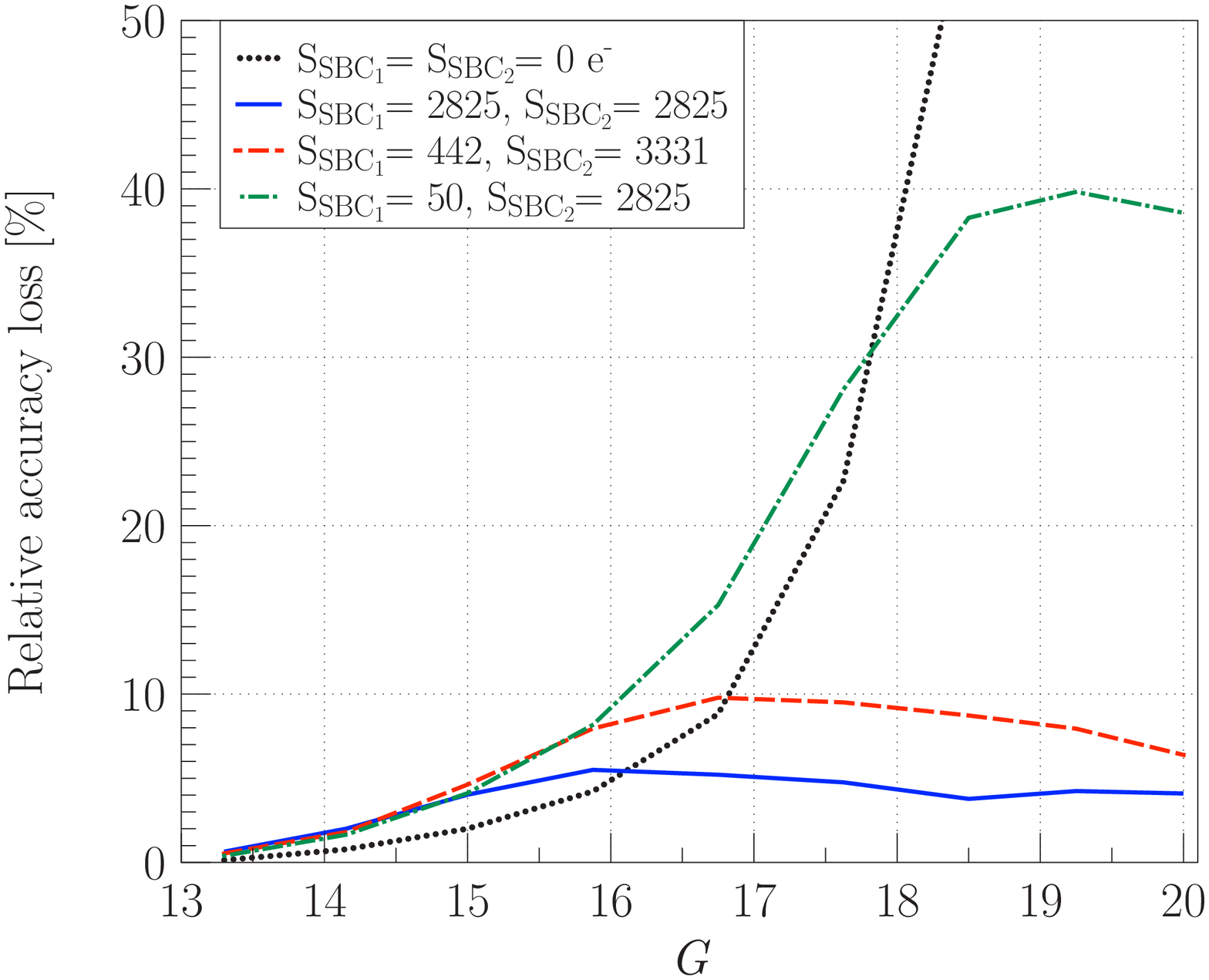}
  \caption{Predicted end-of-life relative image location accuracy loss as a function of magnitude
  ($G$-band) for a 1 s CI delay and five different possible pixel architecture cases.}
  \label{f:mc_cr}
  \includegraphics[width=\columnwidth]{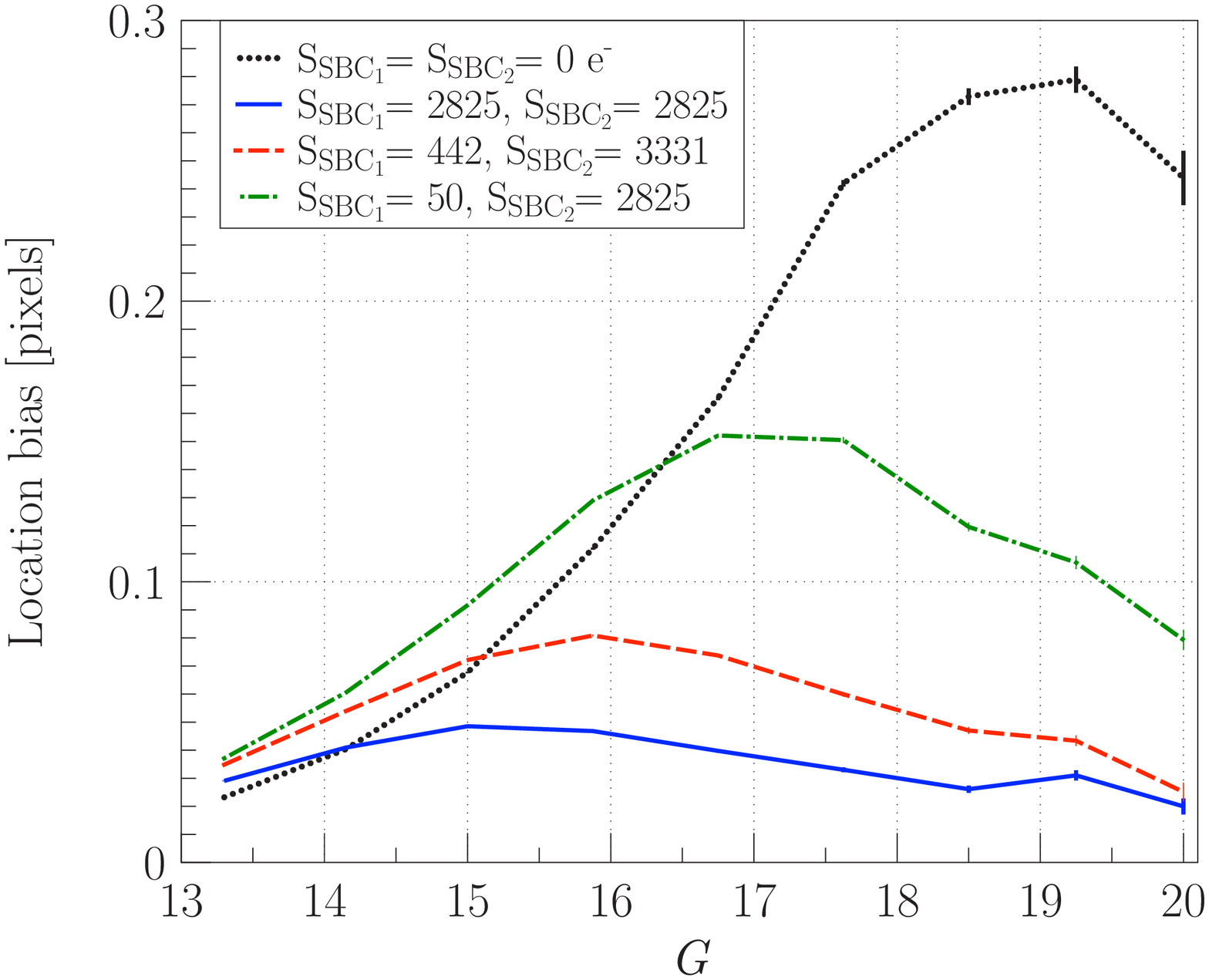}
  \caption{Predicted end-of-life absolute image location bias as a function of magnitude ($G$-band)
  for a CI delay of 1 s and five different pixel architecture cases.  Error bars represent the statistical
  uncertainty, not the standard deviation. Note that here no mitigation procedure at the image processing level was applied.}
  \label{f:mc_bias}
\end{figure}

Fig.~\ref{f:mc_cr} shows the relative loss of accuracy in image location as a function of stellar
magnitude. This is computed by comparing the theoretical limits (the Cram\'{e}r-Rao bounds) to the image location accuracy
 for a damaged and a CTI-free {\gaia}-like image. Zero loss of accuracy corresponds to the CTI-free case \citep[for more details see][]{prodhomme2011b}.
Fig.~\ref{f:mc_bias} shows the image location bias as a function of stellar
magnitude. It is obtained by applying the {\gaia} image location procedure to several hundreds of
profiles without applying any CTI mitigation procedures, the bias itself is computed by subtracting
the true from the estimated location.  Figs.~\ref{f:mc_cr} and \ref{f:mc_bias} provide a comparison
for the five different scenarios described above.  For the simulation we used a DOB that corresponds to the expected average sky brightness of 2.0 \electron pixel$^{-1}$ at readout and
an active trap density of 1 trap pixel$^{-1}$. This particular density was shown
\citep{prodhomme2011b} to reproduce the amplitude of location bias measured using experimental test
data taken 1~s after a CI and for a {\gaia} irradiated CCD with a radiation dose of
$4\times10^9$~protons~cm$^{-2}$ ($10$~MeV equivalent).

From both figures it is clear that the smaller the SBC FWC, the more the location
precision degrades and the greater the bias.  Both  \citet{prodhomme2011b} and \citet{holl2012} cite an early draft of this paper, which at that time only included scenario (i) to show that SBC FWCs with hundreds of electrons in their upper half meet {\gaia}
  requirements: the CTI-induced loss of accuracy of scenario (i) does not exceed $\sim$10 per cent, which is acceptable regarding the {\gaia}
  requirements  and the CTI mitigation
  procedure enables a 90 per cent recovery of the bias, bringing back the residual
  bias of scenario (i) to an acceptable level.

However, with scenario (ii) with SBC FWCs of tens of electrons in their upper half, the bias is a factor of two worse  than scenario (i).  The bias can in principle be calibrated out by using the same mitigation
procedure as foreseen for larger SBC FWCs or the procedure may need to be modified to better model smaller SBC FWCs.  The scenario (ii) loss of accuracy is greater than 10 per cent at $G$ = 16-17 mag and continues to rise as a function of magnitude to a factor of 4-5 worse than scenario (i) at the faint end.  Comparison of Figs. \ref{f:rc_bias} and \ref{f:mc_bias} show that at least one post-2004 CCD (06066-02-02) has a measured bias at the same level as the $(S_{\mathrm{SBC}_1},S_{\mathrm{SBC}_2}) = (2825,2825)$ electrons scenario, which was derived from 03153-15-02 and used by \citet{prodhomme2011b} and \citet{holl2012} to demonstrate that \gaia~meets its astrometric specifications in the presence of radiation damage.  Therefore, it is unlikely that all the CCDs onboard the \gaia~satellite are like scenario (ii).  Even if they are, these simulations are using a radiation dose of $4\times10^9$~protons~cm$^{-2}$ ($10$~MeV equivalent).  The latest predicted end-of-life dose is $3\times10^9$~protons~cm$^{-2}$ ($10$~MeV equivalent).  Therefore, the bias and loss of accuracy are likely to be overestimated.

The location bias of 05256-17-02 in Fig.~\ref{f:rc_bias} is of particular interest because \cite{lindegren2010} carried out a $G \ge 15$ mag astrometric performance analysis of test data acquired using the `sky-like' mask with this particular CCD in which they show that the mission requirements are still met for the faint stars after software radiation damage mitigation has been applied.  In Section \ref{s:pp_results}, we interpreted its SBC FWCs as being smaller than the typical pre-2004 levels i.e. it is not like scenario (i) but its actual values are currently unavailable.  Comparing Figs.~\ref{f:rc_bias} and \ref{f:mc_bias} shows that the shape of the location bias of 05256-17-02 as a function of magnitude is different to the simulated scenario (ii), suggesting that 05256-17-02 is not like scenario (ii) either.  However, we cannot predict how many CCDs onboard \gaia~are like 05256-17-02 so the findings of \cite{lindegren2010} cannot be considered as conclusive proof that \gaia~meets its mission requirements. 

A more representative faint star astrometric performance prediction depends on knowledge of the SBC FWCs onboard the satellite, which is not known.  Therefore, we recommend these should be measured as soon as possible to allow faint star astrometric performance predictions to be updated in case satellite operating conditions or CTI software mitigation can be further optimised to improve the scientific return of \gaia.



%


%
%
\section{Conclusions} \label{s:conc}

ESA's {\gaia} satellite has 106 CCD image sensors which will suffer from increased charge transfer inefficiency (CTI) as a result of radiation damage. 
To aid the mitigation at low signal levels, the CCD design includes Supplementary Buried Channels (SBCs, otherwise known as `notches') within each CCD column.  We present the largest published sample of \gaia~CCD SBC Full Well Capacity (FWC) laboratory measurements and simulations based on 13 devices.  We split this sample into pre- and post-2004 because before manufacturing Flight Models (FMs), e2v changed their photo-lithographic mask set in 2004 but the set was meant to be identical to the one used to manufacture the Demonstrator Models (DMs) and Engineering Models (EMs).  

The pre-2004 sample consists of eight CCDs: one EM, which had its First Pixel Responses (FPR) analysed by \cite{prodhomme2011}, and seven DMs and EMs, which had their FPRs measured by \cite{hopkinson2006}.  We selected a typical and a rare subsample of columns from two of these seven CCDs to model and derive upper and lower half SBC FWCs.  The post-2004 sample consists of five devices: one FM, which was pocket pumped by \cite{kohley2009}, two test structures that we pocket pumped and two EMs, the data from which we re-interpreted in terms of SBC FWCs. 

We find that \gaia~CCDs manufactured post-2004 have SBCs with FWCs in the upper half of each CCD that are systematically smaller by two orders of magnitude ($\le$50 electrons) compared to those manufactured pre-2004 (thousands of electrons).  \gaia's faint star ($13 \le G \le 20$ mag) astrometric performance predictions by \cite{prodhomme2011b} and \cite{holl2012} use pre-2004 SBC FWCs as inputs to their simulations.  However, all the CCDs already integrated onto the satellite for the 2013 launch are post-2004.  SBC FWC measurements are not available for one of our five post-2004 CCDs but the fact it meets \gaia's image location requirements suggests it has SBC FWCs similar to pre-2004.  We are now in a position to answer the three questions posed in the introduction:

\begin{enumerate}
  \item {\it In the absence of testing FM CCD SBC FWCs as a criterion for selecting
    which CCDs should fly on the {\gaia} satellite, is it possible to predict how many will have
    reduced FWCs?}  All the CCDs from the same manufactured batch should nominally have the same SBC FWCs.  However, none of the five post-2004 devices are from the same batch so it is not possible to theoretically predict the SBC FWCs onboard.
  \item {\it Are the SBC FWCs in {\gaia} CCDs used to predict the faint star astrometric performances
    and against which mitigation models are being developed and tested representative of the CCDs that will fly on \gaia?}  A pre-2004 CCD with upper half SBC FWCs of 2825 electrons and the post-2004 CCD without knowledge of its upper half SBC FWCs have been used to predict faint star astrometric performances.  Given that the onboard SBC FWCs are not known, it is not known whether these CCDs are representative of the CCDs that will fly on \gaia~and consequently whether the performance predictions based on these single CCDs are representative of the entire focal plane.
      \item {\it What is the impact of reduced SBC FWCs on the {\gaia} image location accuracy?}  We quantify the impact of reduced SBC FWCs on the {\gaia} image location accuracy, which increases the bias and decreases the accuracy.  However, these numbers are specific to single CCDs, which may not be representative of the entire focal plane.
\end{enumerate}

It is too late to measure the SBC FWCs onboard the satellite before launch.  \gaia's faint star astrometric performance predictions depend on knowledge of the actual SBC FWCs but as these are currently unavailable, it is not known how realistic these predictions are.  Therefore, we suggest \gaia's initial in-orbit calibrations could include a method, which we present in Section \ref{s:recom}, to measure the onboard SBC FWCs.  Realistic faint star astrometric performance predictions at the start of the mission would allow satellite operating conditions or CTI software mitigation to be further optimised to improve the scientific return of \gaia.

It should be emphasized that SBC FWCs were never a formally agreed CCD acceptance criterion in the {\gaia} contract between ESA/EADS-Astrium and e2v, but a design goal. For this reason, it was not systematically tested in the {\gaia} development phase. Assuming extended testing beyond the agreed CCD acceptance criteria would have found the reduced SBC FWCs in post-2004 CCDs, it certainly remains a challenge for CCD testing in large quantities to detect unexpected design and performance issues. While the discovery came too late for {\gaia} to implement design changes, it nevertheless provides a very useful feedback for future CCD developments.  e2v are also developing  the CCDs for ESA's {\it Euclid} satellite (launch 2019), which, despite metrology improvements, will not include SBCs (or anti-blooming drains).

\section{Further work}\label{s:recom}

It is too late to measure the \gaia~SBC FWCs onboard the satellite before launch.  We suggest \gaia's initial in-orbit calibrations (IIOC) could include a method to measure the onboard SBC FWCs.  First Pixel Response (FPR, see Section \ref{s:expt}) measurements will be conducted during the IIOC to measure
radiation trap properties but with only charge injection (CI) at one level.  However, to use FPR to measure SBC FWCs requires many different CI levels (a non-standard procedure), that the CCDs need to have been sufficiently irradiated and that 	the analysis of the data is model dependent.  Pocket pumping (see Section \ref{s:pocket}) cannot be used now that the FM CCDs have been integrated into the satellite because the clock timing in the FM Proximity Electronics Modules does not allow modification for backwards shifts in the image area.  It only works well if the trap density is low, whereas FPR only works well if the trap density is high.  

An alternative to measuring SBC FWCs in a more time-efficient and direct method is the minimum charge injection method (MIM).  MIM injects charge into the first TDI line only i.e. the first pixel in every column (see Appendix for technical details).  Although this single line of charge is transferred through the entire image area of the CCD and read out, it is only designed to be a direct measurement of the SBC FWCs in the first TDI line.  Because of the way e2v manufacture the CCDs (all the SBC doping
implanted simultaneously), intra-stitch block SBC FWCs should be the same.  This paper has shown that the upper half SBC FWCs are more important inputs for simulations than the lower half SBC FWCs.

However, the efficacy of MIM has not been experimentally proven.  This could be because MIM works but has only been tested on CCDs with effectively zero SBC FWCs in their upper halves.  Alternatively, it could be that the measurement method does not work and so returns a null result regardless of SBC FWC. Therefore we plan to test MIM's efficacy on CCDs with large upper half SBC FWCs.  If MIM is experimentally proven, then MIM could be proposed for the IIOC schedule.


\section*{Acknowledgements}

Dr. Gordon Hopkinson (1952-2010) of Surrey Satellite Technology Ltd., Sevenoaks, UK, kindly provided some of the data used in this paper and was very supportive of the project but sadly did not live to see it come to fruition.  This paper is dedicated to his enormous contribution to developing astronomical instrumentation in general and to measuring {\gaia} radiation effects in particular.  We thank David Burt (e2v technologies plc, Chelmsford, UK) for Figs. \ref{f:bil}, \ref{f:bc}, \ref{f:sbc} and \ref{f:pixel} and for his invaluable insight into the design and manufacturing processes of e2v's {\gaia} CCDs through his regular visits to the e2v centre for electronic imaging at The Open University, UK.  We thank EADS-Astrium for their test data taken in Radiation Campaigns 3 and 4.  GMS is funded by the STFC {\gaia} Data Flow System grant and thanks Prof. Mark Cropper for very helpful discussions on the paper.  The work of TP was supported by the European Marie-Curie research training network ELSA (MRTN-CT-2006-033481) and an ESA Research Fellowship.

\begin{table}
  \centering 
  \begin{tabular}{|l|l|}
	\hline
	Acronym 	& Definition	\\
	\hline
	ABD & Anti-Blooming Drain \\
AC & ACross scan \\
AF & Astrometric Field \\
AL & ALong scan \\
BC & Buried Channel \\
BP & Blue Photometer \\
CCD & Charge-Coupled Device \\
CEMGA & Cti Effects Models for GAia \\
CI & Charge Injection \\
CTI & Charge Transfer Inefficiency \\
DM & Demonstration model\\
DOB & Diffuse Optical Background \\
EADS & European Aeronautic Defence and Space company  \\
EM & Engineering Model \\
ESA & European Space Agency \\
FM & Flight Model \\
FPR & First Pixel Response \\
FWC & Full Well Capacity \\
ID & Injection Diode \\
IG & Injection Gate \\
MIM & Minimum Injection Method \\
RC & Radiation Campaign \\
RP & Red Photometer \\
RVS & Radial Velocity Spectrometer \\
SBC & Supplementary Buried Channel \\
TDI & Time-Delayed Integration \\
TDA & Technology Demonstration Activities \\
VTM & Voltage-Tunable Method \\
	\hline
	\end{tabular}
 \caption{List of acronyms used in this paper in alphabetical order.}
\label{t:accronym}
\end{table}


\appendix


\section{Minimum Charge Injection Method} 

\begin{figure}
  \centering
    \includegraphics[width=0.89\columnwidth]{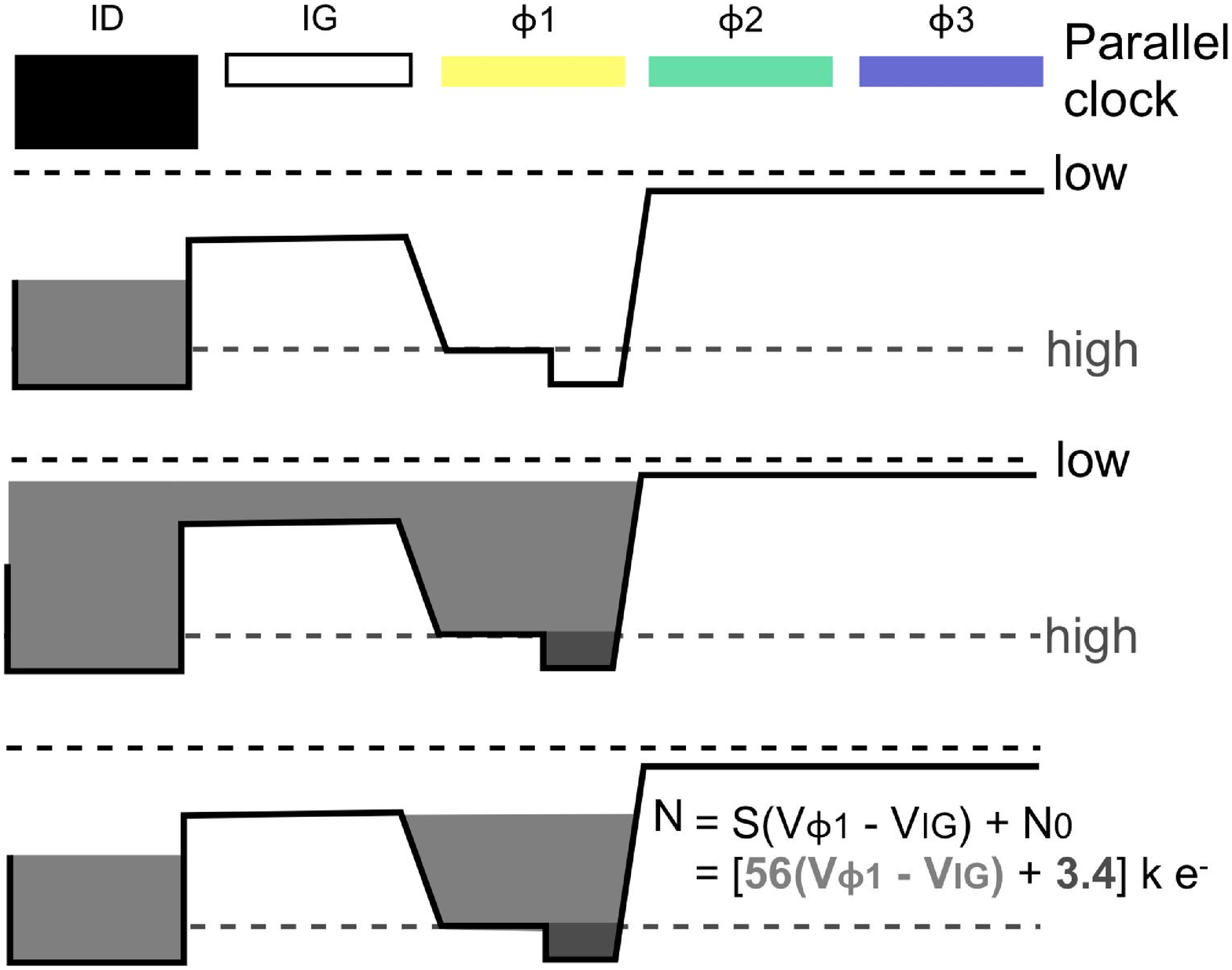}
    \caption{Voltage-tunable method, where the quantity of charge depends on the difference of voltage between
  the `high' level of the image section clock I$\phi$1 and that on IG (Injection Gate). Note that the width of ID (Injection Diode) and IG are only schematic and so actually differ from one another. Also only 3 out of the 4 electrodes contained in a {\gaia} CCD pixel are depicted.
  (\emph{Diagram adapted from \protect\cite{burt2003}.})
 }
  \label{f:vtm}
  \includegraphics[width=0.89\columnwidth]{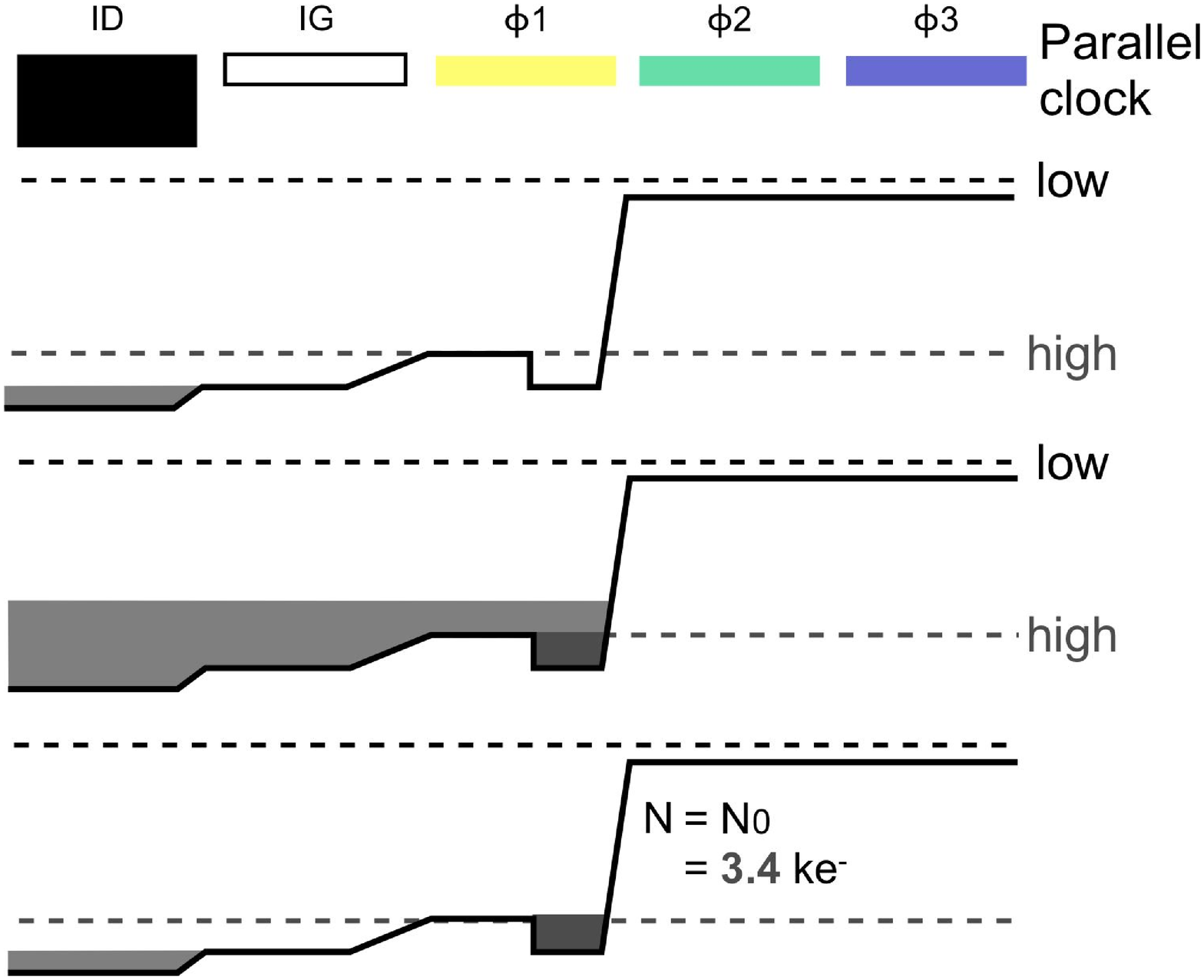}
  \caption{Minimum injection method, where the quantity of charge is independent of the difference of voltage between
  the `high' level of the image section clock I$\phi$1 and that on IG and only depends on the SBC FWC under I$\phi$1 (N$_0$). Note that we did not depict the case for which I$\phi$2 is also
  biased high, such that electrons are also injected under I$\phi$2.  (\emph{Diagram adapted from \protect\cite{burt2003}.})}
  \label{f:mim}
  \end{figure}

Figs.~\ref{f:vtm} and \ref{f:mim} illustrate two
different CI methods available on the {\gaia} CCDs: (top) the Voltage-Tunable Method (VTM)
and (bottom) the Minimum Injection Method (MIM). MIM testing was first suggested by
\cite{holland2004} to investigate continuous, low-level CI. The horizontal lines under the Injection Diode (ID), Injection Gate (IG) and electrodes (I$\phi$1, I$\phi$2 and I$\phi$3) are representative of the potential due to the voltage applied to each.  Like in Figs. \ref{f:bc}, \ref{f:sbc} and \ref{f:pixel}, the convention of plotting low voltages levels as upper horizontal lines and high voltage levels as lower horizontal lines allows the voltage levels to also represent potential levels with electrons filling them up (grey regions) from the bottom (higher voltage) to the top (lower voltage), analogous to water filling up a bath.  The top schematic in Fig.~\ref{f:vtm} shows the IG with a voltage level lower than ID and I$\phi$1 so electrons (like bathwater) cannot flow from ID to I$\phi$1.  The ID fills up with electrons and the middle schematic shows that when the level of electrons is higher than the IG, they can flow to I$\phi$1.  The bottom schematic in Fig.~\ref{f:vtm} shows the ID no longer filling up with electrons.  This stops the flow of electrons across the IG, leaving electrons now occupying both the ID and I$\phi$1. The voltages that are tunable in the VTM are IG and I$\phi$1 (see the equation in Fig.~\ref{f:vtm}).

The idea behind MIM is to transfer charge into I$\phi$1 without the amount of charge stored in I$\phi$1 depending on the voltages applied to IG or I$\phi$1.  This can be achieved by having the potential under ID greater than the potential under IG, which is greater than the potential under I$\phi$1 (see top schematic in Fig.~\ref{f:mim}).  When the ID fills ups with electrons they can flow across the IG into I$\phi$1 (see middle schematic in Fig.~\ref{f:mim}). As with the VTM, when the ID no longer fills up with electrons, the flow of electrons across the IG is stopped.  As the electrons are attracted to the highest voltage, they drain back across IG to ID. However, because there is a SBC under I$\phi$1 and its potential is higher than I$\phi$1, the electrons filling the SBC remain there and only the excess electrons drain out of I$\phi$1.  

In theory MIM enables the precise injection of a very low level of charge independently of the voltage settings. 
The SBC FWC should be more reproducible column-to-column than using only the difference of voltage between the `high' level of I$\phi$1 and that on IG to inject the same small amount of charge. Without a SBC under I$\phi$1, the equations in Fig~\ref{f:vtm} show that to inject 3400~\electron requires a voltage difference of 0.06~V.
Variations in oxide charge from pixel to pixel cause small fixed voltage offsets between columns at this level, which means that a single applied voltage to a single gate results in a non-uniform distribution of potential column-to-column and thus a non-uniform distribution of injected charge column-to-column.
e2v expected MIM to have higher CI uniformity than VTM but this assumes only a small scatter in SBC FWCs.
As explained in Section \ref{s:sbc}, SBC FWC should be uniform within a stitch block but can vary between stitch blocks.
Therefore, MIM may only have higher CI uniformity than VTM within a stitch block but not necessarily over an entire CCD.

The original layout (pre-2004) of the CI structure included I$\phi$1 with the nominal AL dimension
of 3 $\mu$m and an underlying SBC foreshortened in AL by 1 $\mu$m (i.e. AL $\times$ AC = 2 $\times$ 3 $\mu$m).
\cite{burt2005a} reports that e2v MIM testing of this structure did not find N$_{0}$: the electron
FWC under I$\phi$1 (see Fig.~\ref{f:mim}).  This was because the SBC potential was not there as its electrical size was smaller than the geometrical size due to fringing fields and manufacturing tolerances.  Therefore, the CI structure was modified such that
I$\phi$1's AL dimension was extended to 5 $\mu$m and the SBC increased in size to 3 $\times$ 3
$\mu$m to reduce fringing fields and increase the SBC FWC under I$\phi$1 only. 
N$_{0}$ has never been successfully detected.  


%
\bibliographystyle{mn2e}
\bibliography{references}

\end{document}